\documentclass[10pt,prc,showpacs,superscriptaddress,twocolumn]{revtex4}
\usepackage{graphicx}
\begin{document}
\title{Directed and elliptic flow in heavy ion collisions from $E_{\rm
    beam}=90$ MeV/nucleon to $E_{\rm c.m.}=200$ GeV/nucleon}
\author{Hannah Petersen}
\affiliation{Institut f\"ur Theoretische Physik, Johann Wolfgang Goethe-Universit\"at, Max-von-Laue-Str.~1, 
D-60438 Frankfurt am Main, Germany} 
\author{Qingfeng Li}
\affiliation{Frankfurt Institute for Advanced Studies (FIAS), Max-von-Laue-Str.~1, D-60438 Frankfurt am Main,
Germany} 
\author{Xianglei Zhu}
\affiliation{Institut f\"ur Theoretische Physik, Johann Wolfgang Goethe-Universit\"at, Max-von-Laue-Str.~1,
 D-60438 Frankfurt am Main, Germany} 
\affiliation{Frankfurt Institute for Advanced Studies (FIAS), Max-von-Laue-Str.~1, D-60438 Frankfurt am Main,
Germany} 
\affiliation{Physics Department, Tsinghua University, Beijing 100084, China}
\author{Marcus Bleicher}
\affiliation{Institut f\"ur Theoretische Physik, Johann Wolfgang Goethe-Universit\"at, Max-von-Laue-Str.~1,
 D-60438 Frankfurt am Main, Germany}
\begin{abstract}
Recent data from the NA49 experiment on directed and elliptic flow for Pb+Pb reactions at CERN-SPS are compared
 to calculations with a hadron-string transport model, the Ultra-relativistic Quantum Molecular Dynamics (UrQMD) 
model. 
 The rapidity and transverse momentum dependence of the directed and elliptic flow, i.e. $v_{1}$ and $v_2$,  are
 investigated. The flow results are compared to data at three different centrality bins. Generally, a reasonable 
 agreement between the data and the calculations is found.   
Furthermore, the energy excitation functions of $v_1$ and $v_2$
from $E_{\rm beam}=90A$~MeV to $E_{\rm cm}=200A$~GeV are explored
within the UrQMD framework and discussed in the context of the available
data. It is found that, in the energy regime below $E_{\rm beam}\leq 10A$~GeV, the inclusion of nuclear potentials
 is necessary to describe the data. Above $40A$~GeV beam energy, the UrQMD model starts to underestimate the elliptic 
 flow. Around the same energy the slope of the rapidity spectra of the proton directed flow develops negative values. This 
effect is 
 known as the third flow component ("antiflow") and cannot be reproduced by the transport model. These differences can 
 possibly be explained by assuming a phase transition from hadron gas to quark gluon plasma 
 at about $40A$~GeV. 
\end{abstract}
\maketitle
\section{Introduction}

To create extremely hot and dense matter with partons as its
fundamental components - called the Quark-Gluon Plasma (QGP) - is a
major goal of current and future high energy heavy-ion collisions
experiments at SPS, RHIC and LHC \cite{QM2005}. However, due to the
complex nature of the relativistic nucleus-nucleus reactions, the
QGP, if it has been created, escapes direct detection. Therefore, in
order to establish the existence and later on to investigate the
properties of the new state of matter, one must find observables
which allow to deduce the properties of the intermediate (QGP)
state from the final state hadrons. 

The exploration of the transverse collective flow is
the earliest predicted observable to probe heated and 
compressed nuclear matter \cite{Scheid:1974yi}. The transverse flow is intimately connected 
to the pressure gradients. Therefore, it is sensitive to the equation of state (EoS) and might be used to search 
for abnormal matter states and phase transitions \cite{Stoecker:1979mj,Hofmann:1976dy,Stoecker:1986ci}.

The intermediate energy regime available at CERN-SPS or at the future GSI-FAIR facility is often referred to as the 
right place to look for a phase transition to the QGP. Lattice QCD (lQCD) calculations 
\cite{Fodor:2001pe,Fodor:2002km}
 show that the critical temperature is around 170 MeV (for $\mu_b=0$) and the critical energy density is around 
1 $\rm{GeV/fm^3}$. These values can already be reached at 20-30 AGeV beam energy. At finite baryo-chemical potential,
the heated and compressed nuclear matter created at these energies crosses the phase transition line possibly even
on the high $\mu$ side of the critical endpoint. Therefore, it is possible to talk about a phase transition of first 
order, here. During such a first order phase transition the softest point in the equation of state would be mostly 
pronounced. For example, the proton antiflow around midrapidity (``third  flow component''\cite{Csernai:1999nf}) and the 
collapse of the elliptic flow observable have been declared as a signal for the phase transition 
\cite{Stoecker:2004qu,Stoecker:2004xc}.   

In this paper, recent results on proton and pion directed and elliptic flow from the NA49 experiment \cite{Alt:2003ab} 
are investigated and predictions for FAIR are presented. The data is compared to transport model calculations 
(UrQMD v2.2). The proton flow measures the behaviour of the nuclear matter during a heavy ion collision, while the 
flow 
of pions is a sign for the properties of newly produced particles. 

The paper is organized as follows. Section II includes an introduction of the UrQMD model. Section III and IV 
introduce 
the flow systematics and the different measurement methods. Then, in Section V, the directed flow results are shown. 
There are rapidity and transverse momentum distributions for 40 and 160 AGeV beam energy. The centrality dependence 
is also studied. Predictions for rapidity and $p_t$ dependence of $v_1$ for $E_{beam}=20A$~GeV and $30A$~GeV are made. In section V (C) the energy dependence of the slope around midrapidity of the directed flow is investigated in the context of the available data.  
Afterwards, in Section VI, the same analysis for the rapidity and transverse momentum dependence of elliptic flow ($v_2$) is shown. Section VI (C) discusses the 
excitation 
function of elliptic flow over the whole energy range from SIS to RHIC. Section VII summarizes the paper.

\section{The UrQMD model}

For our investigation, the Ultra-relativistic Quantum Molecular 
Dynamics model (UrQMD v2.2)~\cite{Bleicher:1999xi,Bass:1998ca} is applied 
to heavy ion reactions from $E_{\rm beam}= 90A$~MeV to $\sqrt{s_{NN}}=200$ GeV. 
This microscopic transport approach is based on the covariant propagation of
constituent quarks and diquarks accompanied by mesonic and baryonic 
degrees of freedom. It simulates multiple interactions of 
in-going and newly produced particles, the excitation
and fragmentation of colour strings and the formation and decay of
hadronic resonances. 
Towards higher energies, the treatment of sub-hadronic degrees of freedom is
of major importance.
In the present model, these degrees of freedom enter via
the introduction of a formation time for hadrons produced in the 
fragmentation of strings \cite{Andersson:1986gw,Nilsson-Almqvist:1986rx,Sjostrand:1993yb}.
A phase transition to a quark-gluon state is 
not incorporated explicitly into the model dynamics. However, 
a detailed analysis of the model in equilibrium, yields an effective equation of state of 
Hagedorn type \cite{Belkacem:1998gy,Bravina:1999dh}.

The UrQMD transport model is successful in describing the yields and the $p_{t}$ spectra of different particles 
in pp and pA collisions \cite{Bratkovskaya:2004kv}. It has also been applied to study the flow at lower energies 
and at RHIC energies(\cite{Bleicher:2005ti,Winckelmann:1996vu,Bleicher:1998wv,Lu:2006qn,Zhu:2006fb,Zhu:2005qa}). 

\section{Flow systematics}

The first coefficient of the Fourier expansion of the azimuthal distribution of the emitted particles ($v_1$) 
describes 
the directed in-plane flow. The directed flow measures the total amount of transverse flow. It is most pronounced in 
semi-central interactions around target and projectile rapidities where the spectators are deflected away from the 
beam axis due to a bounce-off from the compressed and heated matter in the overlap region. $v_1$ is defined by
\begin{equation}
v_1 \equiv \langle \rm{cos}(\phi-\Phi_{RP})\rangle \quad,
\end{equation}
where $\phi$ denotes the azimuthal angle of one outgoing particles and $\Phi_{RP}$ is the azimuthal angle of the 
reaction plane. The angular brackets denote an average over all considered particles from all events. 

Three different interesting properties of the directed flow have been proposed. 
\begin{itemize}
\item
The time scales probed by the directed flow are set by the crossing time of the Lorentz-contracted
nuclei. Thus, it serves as keyhole to the initial, probably non-equilibrium, stage of the reaction \cite{Kolb:2000sd}.

\item 
The softening of the equation of state in a first order phase transition leads to a decreasing 
directed flow \cite{Rischke:1995pe,Bleicher:2002xx,Soff:1999yg}.

\item
The space-momentum correlation of the emitted particles can be adressed experimentally via the $v_1$ rapidity 
distributions of nucleons and pions.  
\end{itemize}

The second coefficient of the Fourier expansion of the azimuthal distribution of the emitted particles ($v_2$) is 
called 
elliptic flow \cite{Sorge:1998mk,Ollitrault:1992bk,Hung:1994eq,Rischke:1996nq,Sorge:1996pc,Heiselberg:1998es,
Csernai:1999nf,Brachmann:1999xt,Brachmann:1999mp,Zhang:1999rs,Bleicher:2000sx}. This type of flow is strongest around 
central rapidities in semi-peripheral collisions. It is driven by the anisotropy of the pressure {\bf gradients}, due to 
the geometric anisotropy of the initial overlapping region. Therefore, it is a valuable tool to gain insight into the 
expanding stage of the fire ball. $v_2$ is defined by

\begin{equation}
v_2 \equiv \langle \rm{cos}[2(\phi-\Phi_{RP})]\rangle \quad.
\end{equation}

There are two competing effects which lead to contributions with different signs to the integrated $v_2$ value. 
At low energies or early times there is the so called ``squeeze-out''-effect. The spectator matter blocks the emission 
in the impact parameter direction and therefore the flowing matter is ``squeezed''-out perpendicularly to the reaction 
plane. This leads to negative elliptic flow values. The second effect is the so called in-plane flow. This type of 
flow 
becomes important at higher energies and/or later times. At higher bombarding energies ($E_{\rm lab} \ge 10A$~GeV) the 
spectators leave the interaction zone quickly. The remaining hot and dense matter expands almost freely, where the 
surface is such that in-plane emission is preferred. Therefore the elliptic flow receives a positive contribution.

\begin{figure}[hbt]
\centering
\includegraphics[width=10cm]{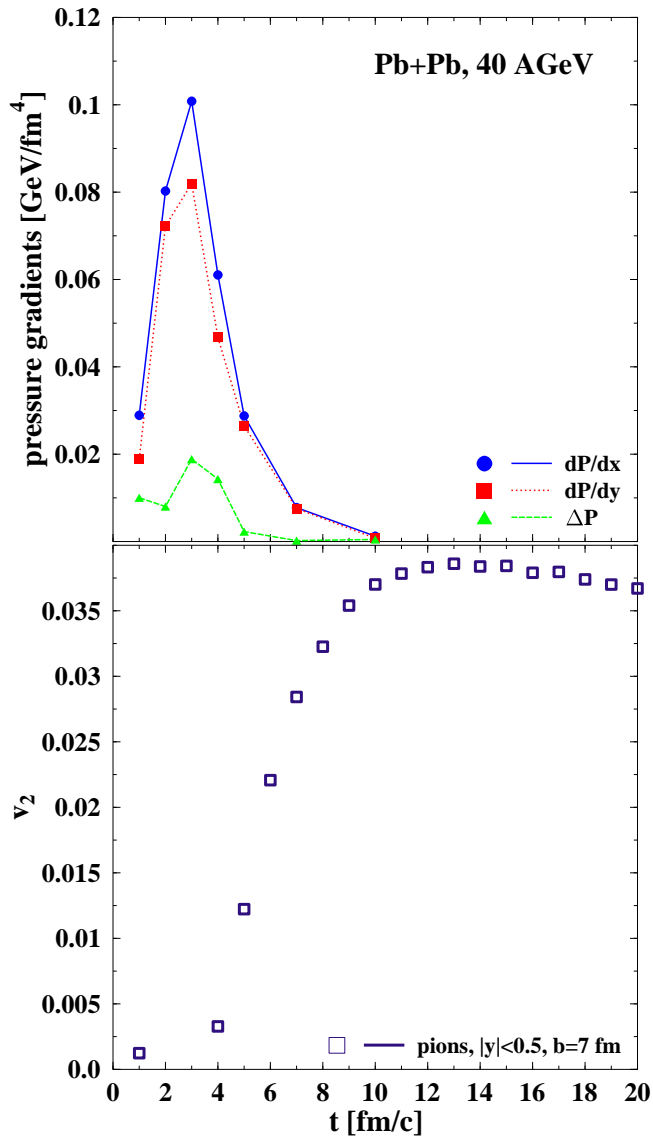}
\caption{(Color online)UrQMD calculation for the time evolution of the pressure gradients and elliptic flow for Pb+Pb interactions at $E_{\rm lab}=40A~$GeV. Top: $dP/dx$ (full line), $dP/dy$ (dotted line) and the difference between these two $\Delta P$ (dashed line) are depicted. Bottom: Elliptic flow of pions (squares) versus time at midrapidity for mid-central collisions (b=7 fm).}
\label{figv2grad40}
\end{figure}

\begin{figure}[hbt]
\centering
\includegraphics[width=10cm]{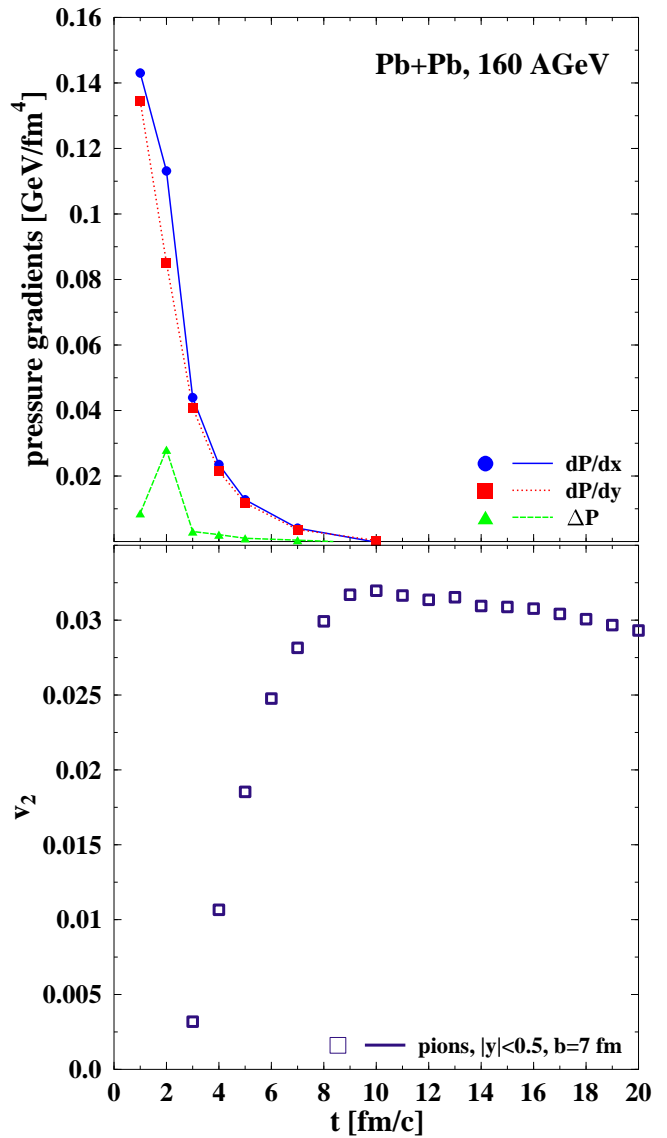}
\caption{(Color online)UrQMD calculation for the time evolution of the pressure gradients and elliptic flow for Pb+Pb interactions at $E_{\rm lab}=160A~$GeV. Top: $dP/dx$ (full line), $dP/dy$ (dotted line) and the difference between these two $\Delta P$ (dashed line) are depicted. Bottom: Elliptic flow of pions (squares) versus time at midrapidity for mid-central collisions (b=7 fm).}
\label{figv2grad160}
\end{figure}

Let us now explore the time evolution of the pressure gradients in connection with the elliptic flow development. The transverse pressure gradients have been calculated for the first 10 fm at $E_{\rm lab}=40A~$GeV (see Fig. \ref{figv2grad40}) and the highest SPS energy (see Fig. \ref{figv2grad160}). In both cases one observes large pressure gradients in the very early stage of the collision. For the lower energy the maximum is reached around $t=3~$fm and for the higher energy it is shifted to even earlier times. The difference between the pressure gradients in x- and y- direction is responsible for the $v_2$ development. As it can be seen in Figs. \ref{figv2grad40}(bottom) and \ref{figv2grad160}(bottom) the temporal evolution of elliptic flow for pions starts exactly after this maximum. The elliptic flow increases during $\sim 6~$fm until it reaches almost its final value. After $t=10~$fm it decreases a little because of resonance decays. So, elliptic flow builds up in the early stage of the collision due to the difference of pressure gradients as it is expected.

\section{Flow measurement methods}

In the UrQMD model calculation of flow, the exact azimuthal angle of the reaction plane $\Phi_{RP}$ is known by 
definition. However, an unambiguous experimental measurement of the azimuthal anisotropic 
flow is not a trivial task due to the unknown orientation of the
reaction plane. Often, experiments use the so called reaction plane
method \cite{Poskanzer:1998yz} to extract the magnitude of flow. In this method, the reaction plane is fixed according 
to the flow vector of the event, then the estimated $v_2$ with
respect to the chosen reaction plane is corrected for the event
plane resolution, which accounts for the error in the deduction of
the reaction plane.

However, these
two-particle correlations based method might suffer from effects which are not
related to the reaction plane, these additional contributions are
usually called non-flow effects \cite{Borghini:2000cm}, such as the overall transverse momentum conservation, small 
angle azimuthal correlations due to final state interactions, resonance decays, jet production \cite{Kovchegov:2002nf} 
and quantum correlations due to the HBT effect \cite{Dinh:1999mn}.  Recently, the
cumulant method was proposed \cite{Borghini:2000sa,Borghini:2001vi} to diminish the non-flow effects. The idea of the 
cumulant method is to extract flow with many-particle cumulants, which are the many-particle correlations with 
subtraction of the contributions from the correlations due to the lower-order multiplets. It is believed that the pure 
many-particle non-flow correlations have much less contributions to the measured flow in the many-particle cumulant 
method.
In other words, the many-particle cumulant method should be much less sensitive to
non-flow effects \cite{Borghini:2000sa,Borghini:2001vi}.

This has been confirmed by the test of the cumulant method with the UrQMD model \cite{Zhu:2006fb,Zhu:2005qa}.
However, the test also shows that, at least for the $v_2$ measurement, the cumulant method is not completely free 
from the effect of event-by-event $v_2$ fluctuations \cite{miller03}. Especially when the genuine $v_2$ signal 
is weak (for example, for the most central events and very peripheral events), the effect of $v_2$ fluctuations could 
be so strong that the results from the cumulant method becomes unreliable.

\section{Directed flow results}
\subsection{Rapidity dependence}

\begin{figure}[hbt]
\centering
\includegraphics[width=10cm]{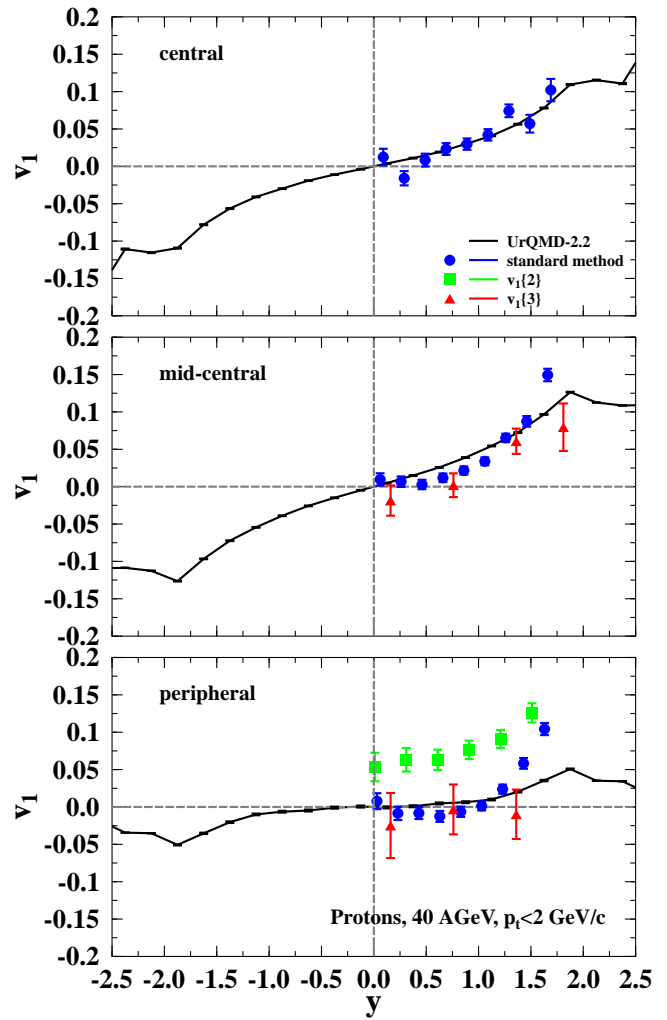}
\caption{(Color online)Directed flow of protons in Pb+Pb collisions at $E_{\rm lab}=40A~$GeV with $p_{t} < 2$~GeV/c. UrQMD 
calculations are depicted with black lines. The symbols are NA49 data from different analysis methods. The standard 
method (circles), cumulant method of order 2 (squares) and cumulant method of order 3 (triangles) are depicted. The 
12.5\% most central collisions are labeled as central, the centrality 12.5\% -33.5\% as mid-central and 33.5\% -100\% 
as peripheral. For the model calculations the corresponding impact parameters of $b \le 3.4$~fm for central, 
$b=5-9$~fm 
for mid-central and $b= 9-15$~fm for peripheral collisions have been used.}
\label{figv1yp40}
\end{figure}

Fig. \ref{figv1yp40} shows the rapidity dependence of the directed flow of protons for central ($b \le 3.4$ fm), 
mid-central ($b=5-9$ fm) and peripheral ($b \ge 9$ fm) Pb+Pb reactions at $E_{\rm lab}=40A$~GeV. The symbols denote 
data
by the NA49 collaboration analyzed with different methods \cite{Alt:2003ab}. UrQMD calculations are depicted with 
black 
lines. The directed flow is most pronounced at high rapidity values where the bounced off spectator matter sits.

The strong centrality dependence of $v_1$ can be seen, as the shape of the curves changes drastically from central 
to peripheral collisions. Overall, the model calculations are in line with the data from the event plane method 
(standard method, full circle). It is interesting to note that one kind of non-flow effects, i.e. momentum 
conservation, has been subtracted in this reaction plane method already \cite{Borghini:2002mv}. Unfortunately, 
the 2-particle cumulant measurements seem to be affected by the non-flow effects which are in this analysis dominated 
by momentum conservation. This effect is most pronounced for peripheral collisions. Here, the experimental data points 
do not approach zero at midrapidity. For the more reliable 3-particle cumulant method \cite{Borghini:2002vp}, the 
experimental results agree well with those of the reaction plane method within the statistical error (please see 
also the following $v_1$ data). This indicates that the momentum conservation corrected reaction plane method gives 
also reliable $v_1$ data.

\begin{figure}[hbt]
\centering
\includegraphics[width=10cm]{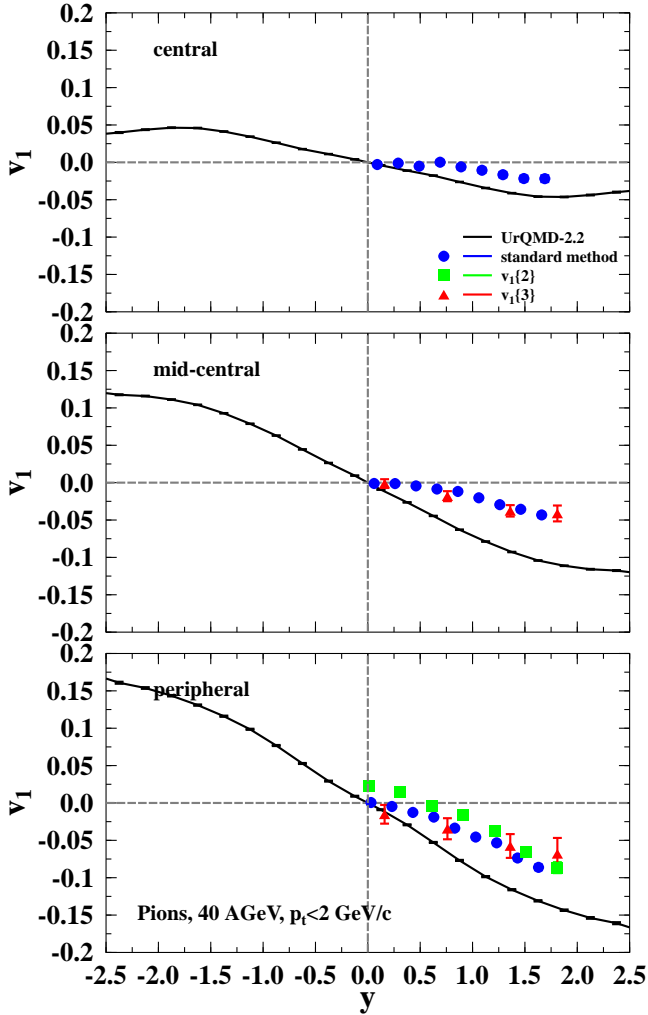}
\caption{(Color online)Directed flow of pions in Pb+Pb collisions at $E_{\rm lab}=40A~$GeV with $p_{t} < 2$~GeV/c. UrQMD 
calculations are depicted with black lines. The symbols are NA49 data from different analysis methods. The standard 
method (circles), cumulant method of order 2 (squares) and cumulant method of order 3 (triangles) are depicted. The 
12.5\% most central collisions are labeled as central, the centrality 12.5\% -33.5\% as mid-central and 33.5\% -100\% 
as peripheral. For the model calculations the corresponding impact parameters of $b \le 3.4$~fm for central, 
$b=5-9$~fm
 for mid-central and $b= 9-15$~fm for peripheral collisions have been used.}
\label{figv1ypi40}
\end{figure}

Fig. \ref{figv1ypi40} shows directed flow of pions at $E_{\rm
lab}=40A$~GeV. The different sign with respect to the proton flow can
be explained by shadowing. The pions are newly produced mesons and therefore they are composed by a quark and 
especially an anti-quark. Therefore, the cross-section of the pions with the nuclear matter is so large that they
cannot escape in the direction where the rest of the colliding nuclei/spectator matter resides.

 For peripheral collisions the measured directed
flow reaches the same amount as for protons (about 10\% at $y\approx 2$), but
for central collisions it is much less (only about 2.5\% at $y \approx 2$). The
UrQMD calculations overestimate the pion directed flow at large rapidities by about a factor
of two for all centralities. This overestimation might be explainable if one assumes that the NA49
collaboration is not able to measure all produced pions. Especially pions produced from nucleons which fly straight 
ahead through the collision producing only one
or two pions appear only in the veto calorimeter. However, in the model calculations every produced pion within
the given rapidity and transverse momentum bin is taken into account.

\begin{figure}[hbt]
\centering
\includegraphics[width=10cm]{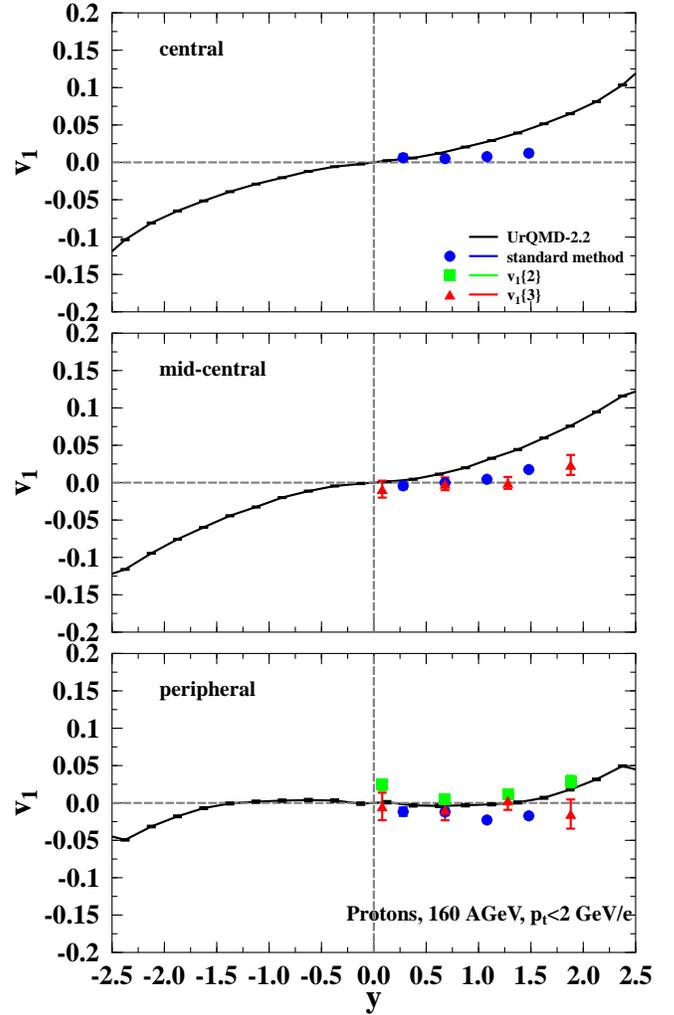}
\caption{(Color online)Directed flow of protons in Pb+Pb collisions at $E_{\rm lab}=160A~$GeV with $p_{t} < 2$~GeV/c. UrQMD 
calculations are depicted with black lines. The symbols are NA49 data from different analysis methods. The standard 
method (circles), cumulant method of order 2 (squares) and cumulant method of order 3 (triangles) are depicted. The 
12.5\% most central collisions are labeled as central, the centrality 12.5\% -33.5\% as mid-central and 33.5\% -100\% 
as peripheral. For the model calculations the corresponding impact parameters of $b \le 3.4$~fm for central, 
$b=5-9$~fm 
for mid-central and $b= 9-15$~fm for peripheral collisions have been used.}
\label{figv1yp160}
\end{figure}
In Fig. \ref{figv1yp160}, directed flow of protons at $E_{\rm
lab}=160A$~GeV is shown. In this case the model calculations slightly
overestimate the flow at higher rapidities in contrast to the proton
flow at $E_{\rm lab}=40A$~GeV. Surprisingly, the data stays almost constant at
zero. Even in peripheral collisions there are only about $v_1=2.5\%$ at
$y=1.5$ for the reaction plane method analysis. 
The flat shape of the curve with respect to the lower energy results can be
explained by the spectators sit at higher rapidity values in this
plot. This spectator matter is responsible for the directed flow near beam/target rapidities. Since there are no 
data points of the reaction plane method
above $y=1.5$ this increase is not seen.

\begin{figure}[hbt]
\centering
\includegraphics[width=10cm]{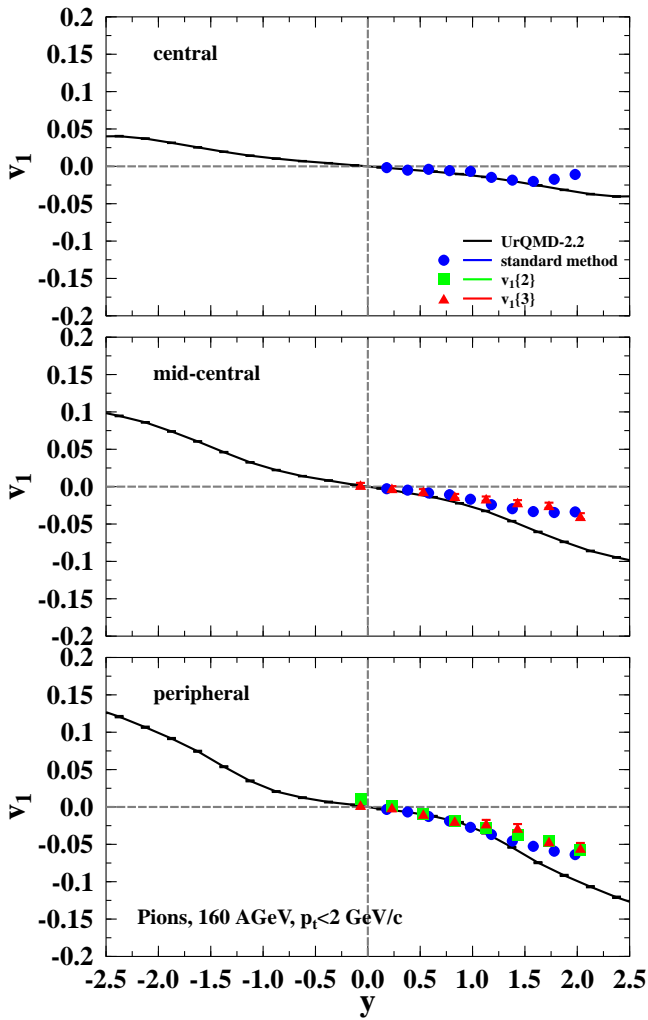}
\caption{(Color online)Directed flow of pions in Pb+Pb collisions at $E_{\rm lab}=160A~$GeV with $p_{t} < 2$~GeV/c. UrQMD 
calculations are depicted with black lines. The symbols are NA49 data from different analysis methods. The standard 
method (circles), cumulant method of order 2 (squares) and cumulant method of order 3 (triangles) are depicted. The 
12.5\% most central collisions are labeled as central, the centrality 12.5\% -33.5\% as mid-central and 33.5\% -100\% 
as peripheral. For the model calculations the corresponding impact parameters of $b \le 3.4$~fm for central, 
$b=5-9$~fm 
for mid-central and $b= 9-15$~fm for peripheral collisions have been used.}
\label{figv1ypi160}
\end{figure}

For the pion directed flow at the higher SPS energy (Fig.
\ref{figv1ypi160}), one observes a reasonable agreement between the model calculations
and the data from all three measurement methods, i.e. reaction plane method, $v_1\{2\}$ and $v_1\{3\}$. Above $y=1$, 
the UrQMD results increase slightly stronger than the data. The centrality
dependence is reproduced correctly. Both in the model and in the
experimental data there is an increase in flow for more peripheral
collisions.

\begin{figure}[hbt]
\centering
\includegraphics[width=10cm]{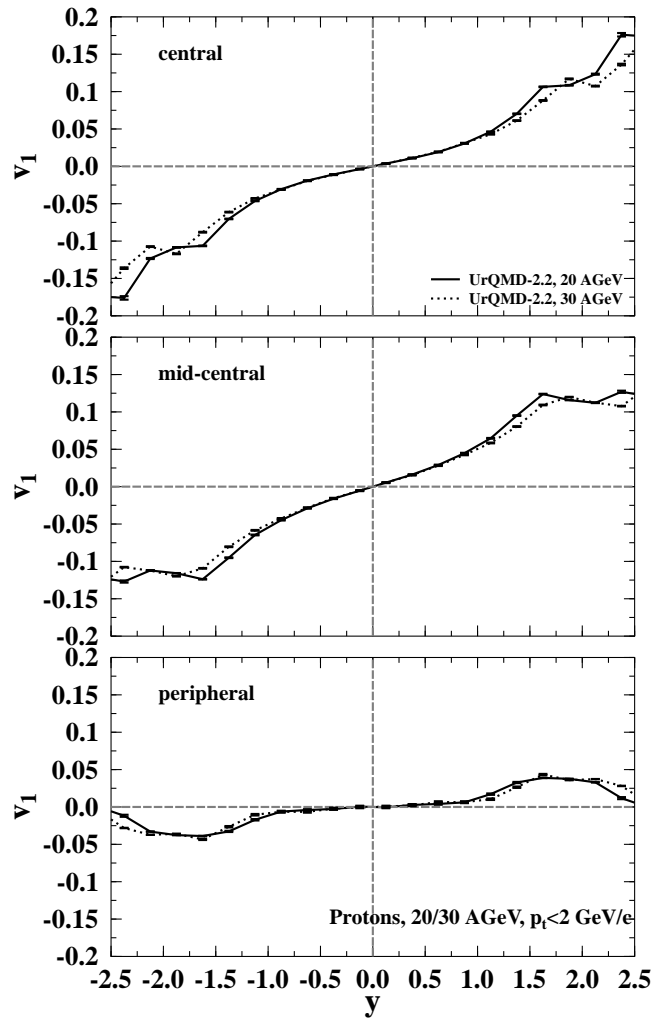}
\caption{Directed flow of protons in Pb+Pb collisions at $E_{\rm lab}=20A~$GeV and $E_{\rm lab}=30A~$GeV with 
$p_t < 2$~GeV/c. UrQMD calculations for 20 AGeV are depicted with solid lines while the results for 30 AGeV are 
depicted by dashed lines. Impact parameters of $b \le 3.4$~fm for central, $b=5-9$~fm for mid-central and 
$b= 9-15$~fm for peripheral collisions have been used.}
\label{figv1yp20}
\end{figure}
\begin{figure}[hbt]
\centering
\includegraphics[width=10cm]{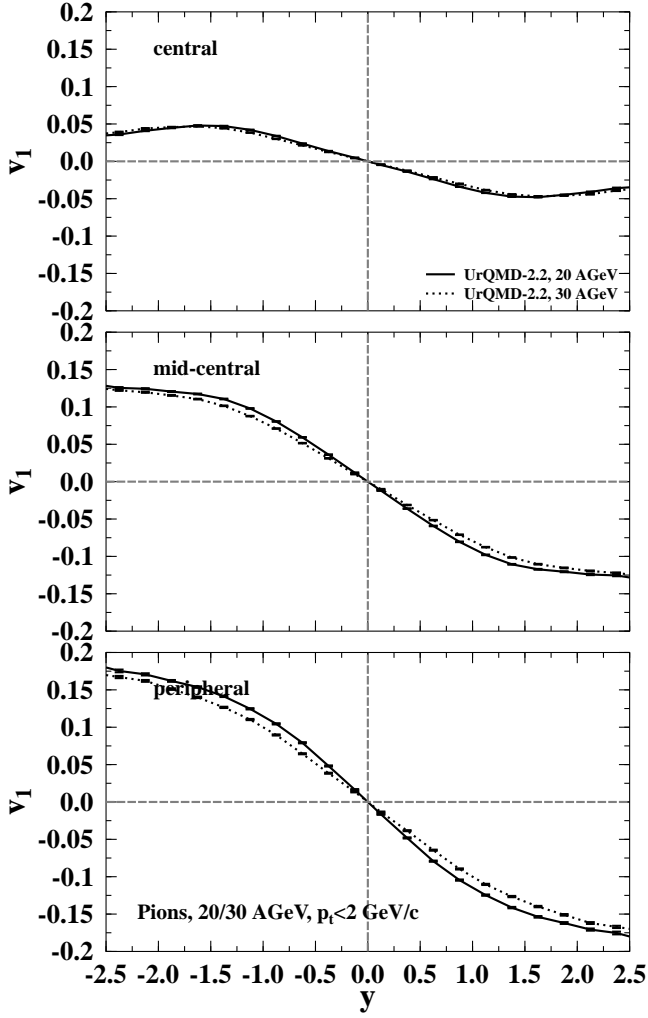}
\caption{Directed flow of pions in Pb+Pb collisions at $E_{\rm lab}=20A~$GeV and $E_{\rm lab}=30A~$GeV with 
$p_t < 2$~GeV/c. UrQMD calculations for 20 AGeV are depicted with solid lines while the results for 30 AGeV are 
depicted by dashed lines. Impact parameters of $b \le 3.4$~fm for central, $b=5-9$~fm for mid-central and 
$b= 9-15$~fm for peripheral collisions have been used.}
\label{figv1ypi20}
\end{figure}
Fig. \ref{figv1yp20} and Fig. \ref{figv1ypi20} are predictions for
the rapidity dependence of proton and pion directed flow at $E_{\rm lab}=20A$~GeV and $E_{\rm lab}=30A$~GeV. The shown
results are calculated using the UrQMD model with a transverse momentum
cut of $p_t < 2~$GeV/c. In the present model, the calculated directed flow results at these energies under 
investigation at the
new FAIR facility at GSI look rather similar to that at $E_{\rm
lab}=40A$~GeV. There is also an inverse centrality dependence for protons and pions in the amount of the directed 
flow. 
It is very interesting to see if the experimental proton flow data at this energy will show the negative slope around 
midrapidity as predicted.

\subsection{Transverse momentum dependence}

\begin{figure}[hbt]
\centering
\includegraphics[width=10cm]{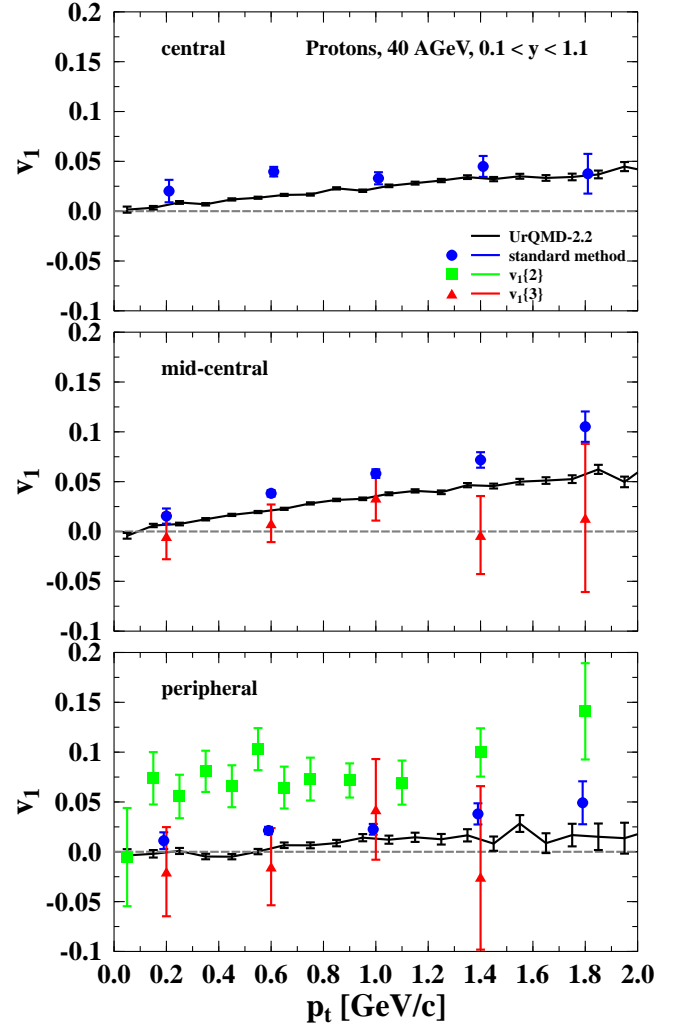}
\caption{(Color online)Directed flow of protons in Pb+Pb collisions at $E_{\rm lab}=40A~$GeV with $0.1 < y < 1.1$. UrQMD 
calculations 
are depicted with black lines. The symbols are NA49 data from different analysis methods. The standard method 
(circles), 
cumulant method of order 2 (squares) and cumulant method of order 3 (triangles) are depicted. The 12.5\% most central 
collisions are labeled as central, the centrality 12.5\% -33.5\% as mid-central and 33.5\% -100\% as peripheral. For 
the model calculations the corresponding impact parameters of $b \le 3.4$~fm for central, $b=5-9$~fm for mid-central 
and $b= 9-15$~fm for peripheral collisions have been used.}
\label{figv1ptp40}
\end{figure}

Fig. \ref{figv1ptp40} shows directed flow of protons at $E_{\rm
lab}=40A$~GeV as a function of transverse momentum. For the calculation the same
rapidity cut ($0.1 < y < 1.1$) as in the data obtained with the cumulant method has been used. For the
standard reaction plane analysis the different cut of $0.1 < y <1.8$ has been applied.
The different cuts have been used by the experimental collaboration to improve the statistics and reduce the 
time amount for the analysis. Since the higher order
cumulant method measurements are the most reliable way to reduce systematic
errors \cite{Zhu:2006fb,Zhu:2005qa}, the comparisons are done with the cuts for the cumulant method. 
The calculations are in line with the $v_1\{3\}$
data for mid-central and peripheral collisions. Unfortunately, the
statistical errors are the biggest for this method and there are only a few data
points because it is the newest and most elaborate analysis. 

\begin{figure}[hbt]
\centering
\includegraphics[width=10cm]{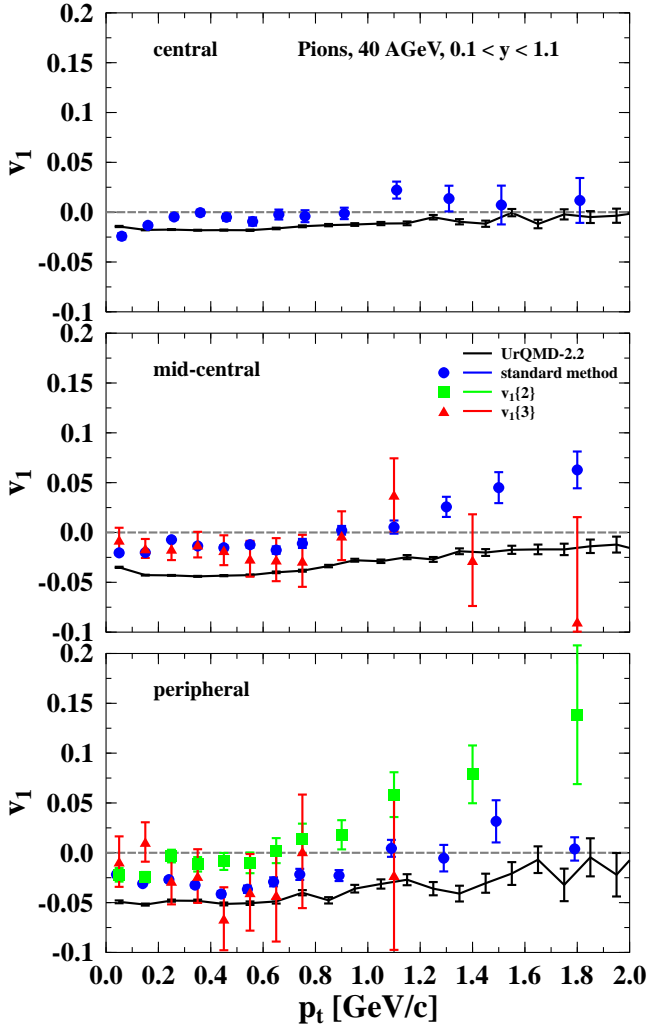}
\caption{(Color online)Directed flow of pions in Pb+Pb collisions at $E_{\rm lab}=40A~$GeV with $0.1 < y < 1.1$. UrQMD calculations 
are depicted with black lines. The symbols are NA49 data from different analysis methods. The standard method 
(circles), 
cumulant method of order 2 (squares) and cumulant method of order 3 (triangles) are depicted. The 12.5\% most central 
collisions are labeled as central, the centrality 12.5\% -33.5\% as mid-central and 33.5\% -100\% as peripheral. For 
the model calculations the corresponding impact parameters of $b \le 3.4$~fm for central, $b=5-9$~fm for mid-central 
and $b= 9-15$~fm for peripheral collisions have been used.}
\label{figv1ptpi40}
\end{figure}

The transverse momentum dependence of the pion directed flow at $E_{\rm lab}=40A$~GeV (Fig. \ref{figv1ptpi40}) shows 
many uncertainties. The data differs very much depending on the analysis
method. Furthermore, the cumulant method of order 3 has large
statistical error bars. The UrQMD calculations are negative as the sign convention suggests. It is remarkable that 
the directed flow does not show any clear transverse momentum dependence.

\begin{figure}[hbt]
\centering
\includegraphics[width=10cm]{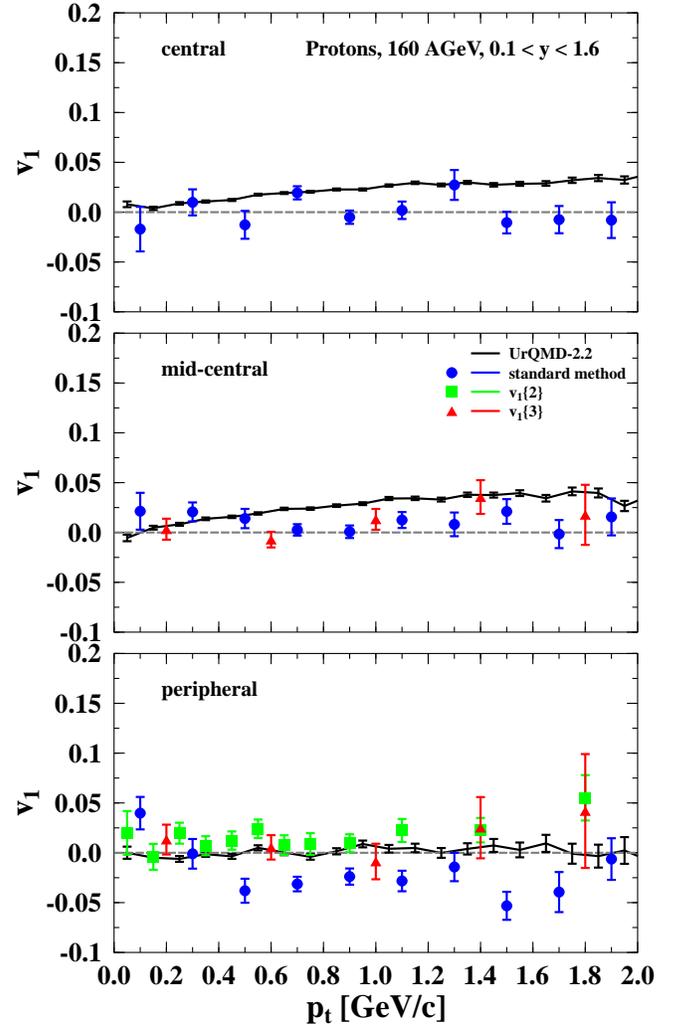}
\caption{(Color online)Directed flow of protons in Pb+Pb collisions at $E_{\rm lab}=160A~$GeV with $0.1 < y < 1.6$. UrQMD 
calculations are depicted with black lines. The symbols are NA49 data from different analysis methods. The standard 
method (circles), cumulant method of order 2 (squares) and cumulant method of order 3 (triangles) are depicted. The 
12.5\% most central collisions are labeled as central, the centrality 12.5\% -33.5\% as mid-central and 33.5\% -100\% 
as peripheral. For the model calculations the corresponding impact parameters of $b \le 3.4$~fm for central, 
$b=5-9$~fm 
for mid-central and $b= 9-15$~fm for peripheral collisions have been used.}
\label{figv1ptp160}
\end{figure}

Next, we turn to the results for $160A$~GeV collisions. For central collisions the calculated transverse momentum 
dependence of the directed flow of protons at $E_{\rm lab}=160A$~GeV (see Fig. \ref{figv1ptp160}) starts at
zero for $p_t=0$ and increases steadily until 2\% at $p_t=2~$GeV/c. The
measured data fluctuates between $+2\%$ and $-2\%$. Going to mid-central
collisions the calculations are quite in line with the third order
cumulant method and the standard method data. In
peripheral collisions there are big systematic uncertainties because
especially around $p_t=1.8~$GeV/c the data points differ between $-4\%$
and $+5.5\%$. The second order cumulant
results are higher, but they are not corrected due to momentum conservation. The negative directed flow of protons 
is consistent with the rapidity distribution. Because of the negative slope around midrapidity at the higher 
SPS energy the protons fly in the ``wrong'' direction.  

\begin{figure}[hbt]
\centering
\includegraphics[width=10cm]{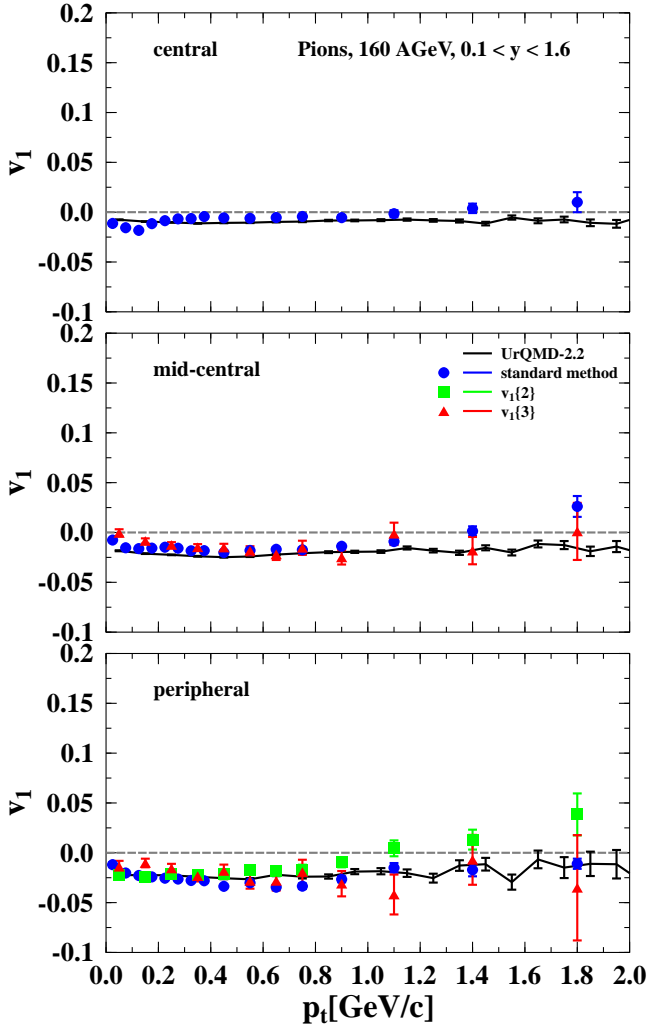}
\caption{(Color online)Directed flow of pions in Pb+Pb collisions at $E_{\rm lab}=160A~$GeV with $0.1 < y < 1.6$. UrQMD calculations 
are depicted with black lines. The symbols are NA49 data from different analysis methods. The standard method 
(circles), 
cumulant method of order 2 (squares) and cumulant method of order 3 (triangles) are depicted. The 12.5\% most central 
collisions are labeled as central, the centrality 12.5\% -33.5\% as mid-central and 33.5\% -100\% as peripheral. For 
the model calculations the corresponding impact parameters of $b \le 3.4$~fm for central, $b=5-9$~fm for mid-central 
and $b= 9-15$~fm for peripheral collisions have been used.}
\label{figv1ptpi160}
\end{figure}

The calculated directed flow for pions at $E_{\rm lab}=160A$~GeV, as depicted in Fig. \ref{figv1ptpi160}, stays
constant as a function of the transverse momentum at $(-1)\%$ for central, at $(-2)\%$ for mid-central and at $(-3)\%$
for peripheral collisions. In mid-central and peripheral collisions this
trend is in line with the third order cumulant measurement. Especially
the data from second order cumulant increase up to $4\%-6\%$. This is probably
due to momentum conservation which is not taken into account to correct
the data. The negative directed flow of pions is consistent with the
rapidity dependence in the chosen bin.

\begin{figure}[hbt]
\centering
\includegraphics[width=10cm]{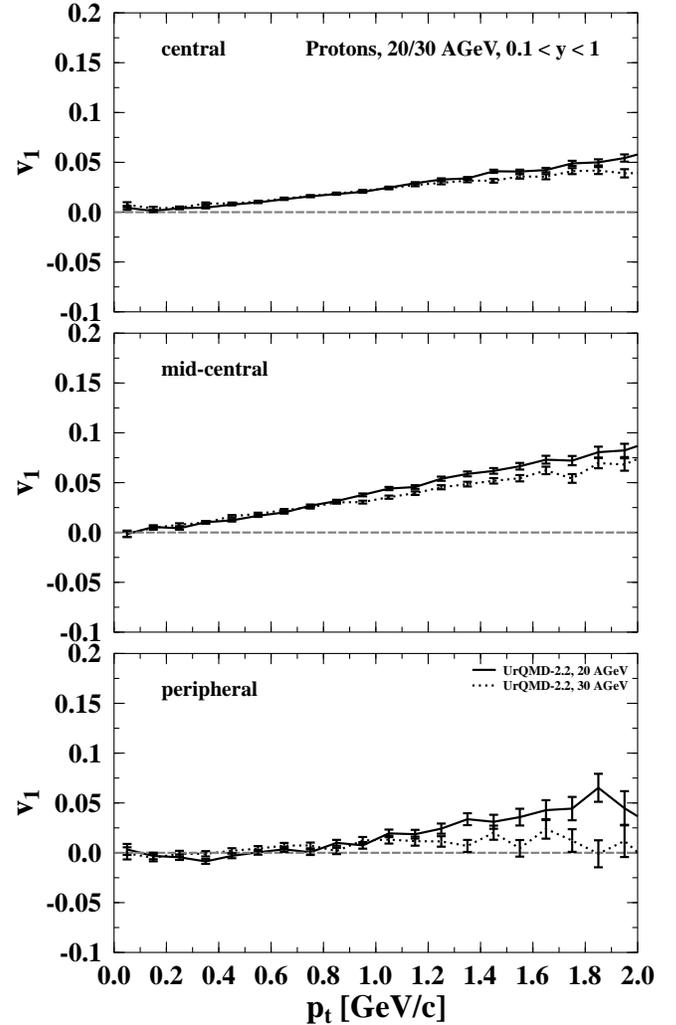}
\caption{Directed flow of protons in Pb+Pb collisions at $E_{\rm lab}=20A~$GeV and $E_{\rm lab}=30A~$GeV with 
$0.1 < y < 1$. UrQMD calculations for 20 AGeV are depicted with solid lines while the results for 30 AGeV are depicted 
by dashed lines. Impact parameters of $b \le 3.4$~fm for central, $b=5-9$~fm for mid-central and $b= 9-15$~fm for 
peripheral collisions have been used.}
\label{figv1ptp20}
\end{figure}
\begin{figure}[hbt]
\centering
\includegraphics[width=10cm]{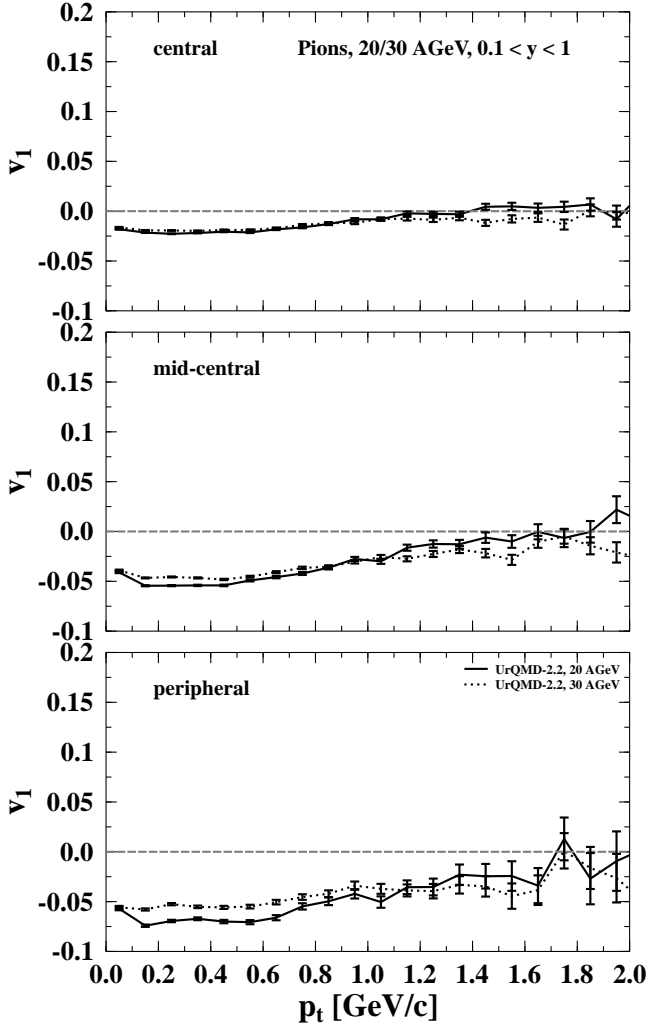}
\caption{Directed flow of pions in Pb+Pb collisions at $E_{\rm lab}=20A~$GeV and $E_{\rm lab}=30A~$GeV with 
$0.1 < y < 1$. UrQMD calculations for 20 AGeV are depicted with solid lines while the results for 30 AGeV are 
depicted by dashed lines. Impact parameters of $b \le 3.4$~fm for central, $b=5-9$~fm for mid-central and 
$b= 9-15$~fm for peripheral collisions have been used.}
\label{figv1ptpi20}
\end{figure}

Finally, we show the prediction for the transverse momentum dependence of directed flow
at $E_{\rm lab}=20A$~GeV and $E_{\rm lab}=30A$~GeV in Figs. \ref{figv1ptp20} and \ref{figv1ptpi20}. In the present 
calculations, the shape and magnitude of the flow are similar to the results at $40A$~GeV. The directed flow of pions 
at $20A$~GeV 
and $30A$~GeV looks also rather similar to the
calculations for the pion directed flow at $40A$~GeV. But there is a
difference for peripheral collisions. The $v_1(p_t)$ value is only about $(-2.5) \%$ compared 
to $(-5)\%$ at $E_{\rm lab}=40A$~GeV.

\subsection{Excitation function}
\begin{figure}[hbt]
\centering
\includegraphics[width=10cm]{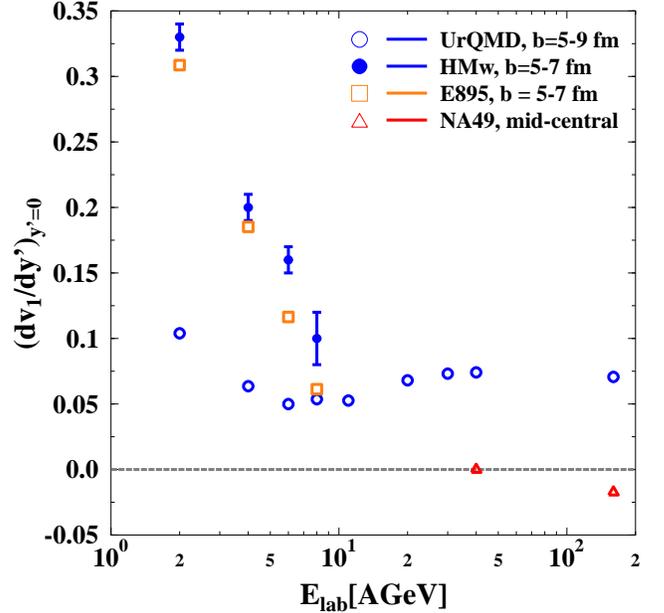}
\caption{(Color online)Slope of $v_1(y)$ of protons around midrapidity extracted from normalized ($y^\prime = y/y_b$) rapidity distributions. The data are taken 
from
 E895 (squares)\cite{Liu:2000am} and NA49 (triangles)\cite{Alt:2003ab}. UrQMD calculations with included mean field 
(HMw)are 
 depicted with full circles. Open circles depict UrQMD calculation in the cascade mode.}
\label{fig_slope}
\end{figure}

To characterize the amount and the direction of the directed flow of protons over the energy range from $2-160~A$GeV one can 
extract the slope around midrapidity from the normalized rapidity distributions usually referred to as the 
``F'' parameter \cite{Liu:2000am}. ``Normalized'' means in this case $y/y_b$ where $y_b$ is the beam rapidity. This 
normalization accounts for the trivial energy dependence of the slope. The values for the slope in Fig. 
\ref{fig_slope} 
have been extracted via a polynomial fit of the form $a \cdot x + b \cdot x^3$ with $x=y/y_b$. At low energies, one 
observes that the inclusion of a nuclear potential is needed to reproduce the data. Here is shown the calculation with 
included mean field from a hard equation of state with momentum dependence and medium-modified nucleon-nucleon cross 
sections (HMw)\cite{Li:2005gf,Li:2006ez}. At higher energies the calculation has been performed in the cacsade mode
without the additional inclusion of nuclear potentials. 

At SPS energies the data develops even negative values for the slope around midrapidity 
\cite{Stoecker:2004qu,Stoecker:2004xc}. This behaviour cannot be reproduced within the transport model calculation. 
However, ideal hydro calculations have predicted the appearance of a
so-called ''third flow component'' \cite{Csernai:1999nf} or ''antiflow''
\cite{Brach00} at finite impact parameters. In these analysis' it was pointed out that this ``antiflow'' develops if
the matter undergoes a first order phase transition to the QGP. In contrast, a hadronic EoS without
QGP phase transition did not yield such an exotic ''antiflow''
(negative slope) wiggle in the proton flow $v_1(y)$ at low energies.

\section{Elliptic flow}
\subsection{Rapidity dependence}
\begin{figure}[ht]
\centering
\includegraphics[width=10cm]{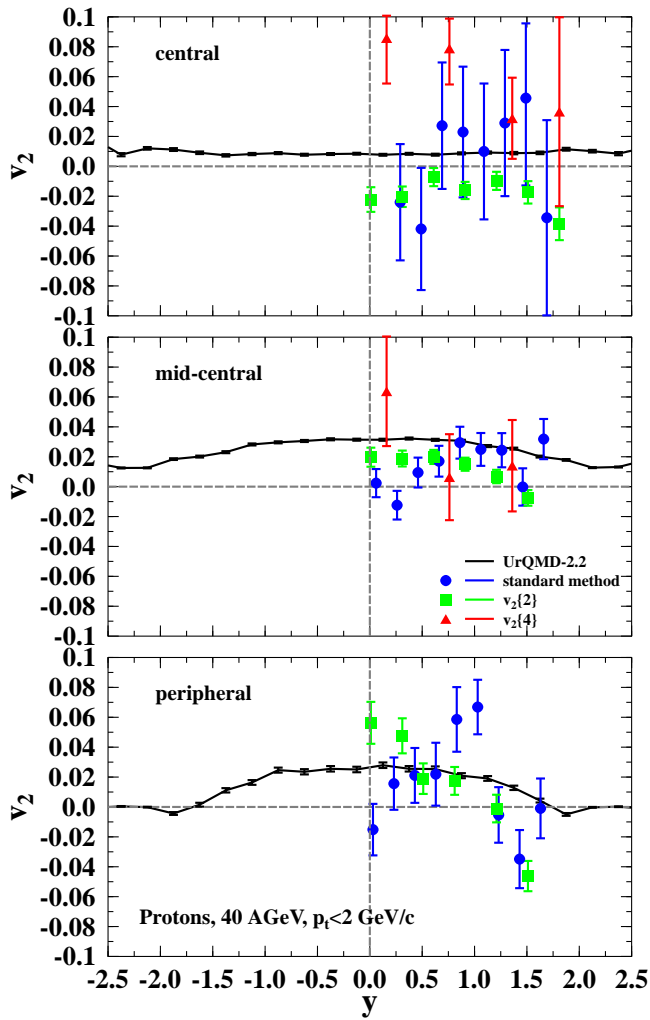}
\caption{(Color online)Elliptic flow of protons in Pb+Pb collisions at $E_{\rm lab}=40A~$GeV with $p_{t} < 2$~GeV/c. UrQMD 
calculations are depicted with black lines. The symbols are NA49 data from different analysis methods. The standard 
method (circles), cumulant method of order 2 (squares) and cumulant method of order 3 (triangles) are depicted. 
The 12.5\% most central collisions are labeled as central, the centrality 12.5\% -33.5\% as mid-central and 
33.5\% -100\% as peripheral. For the model calculations the corresponding impact parameters of $b \le 3.4$~fm for 
central, $b=5-9$~fm for mid-central and $b= 9-15$~fm for peripheral collisions have been used.}
\label{figv2yp40}
\end{figure}

Elliptic flow develops because of the almond shape of the overlapping region
in a heavy ion collision. The breakdown of proton elliptic flow at
$E_{\rm lab}=40A$~GeV has been stressed as a signal for the observation
of a first order phase transition \cite{Stoecker:2004xc}. As can be seen from the NA49 data in Fig. \ref{figv2yp40}, 
the elliptic flow parameter
$v_2$ vanishes at midrapidity only for the standard reaction plane method data, which could be affected by the 
non-flow effects. The cumulant
measurements show a completely different shape. The second and fourth
order $v_2$ measurements for peripheral and mid-central collisions increase to about
$6-8\%$ around midrapidity not consistent with zero. Looking now at the UrQMD
calculations one cannot observe an overestimation, but finds
that the results are compatible with the data. Therefore, before one can draw concise conclusions the systematic
and statistical uncertainties of the data must be resolved.

\begin{figure}[t]
\centering
\includegraphics[width=10cm]{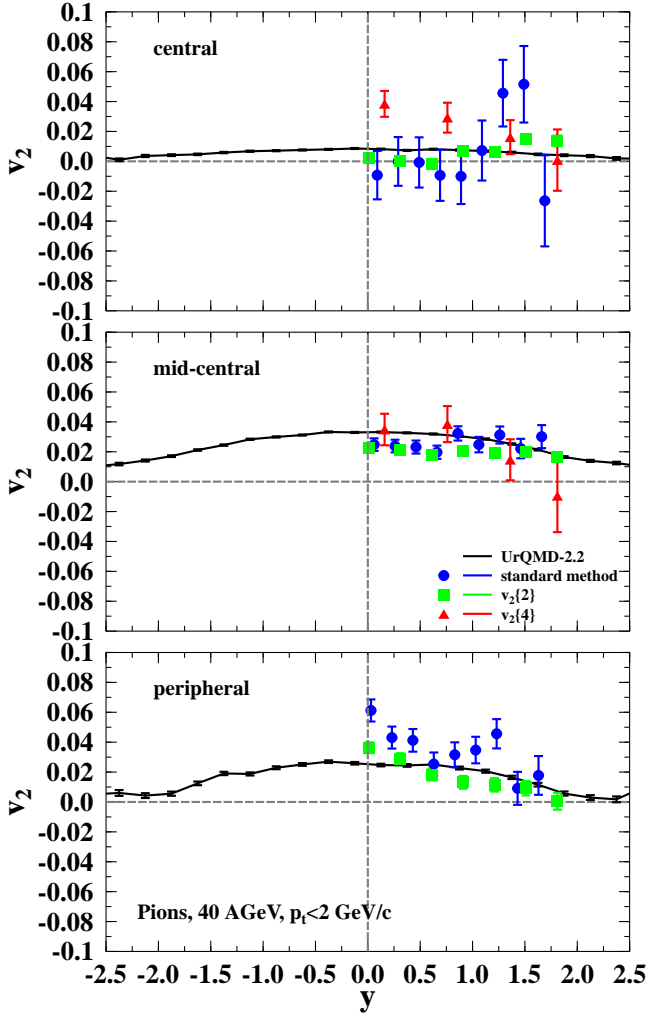}
\caption{(Color online)Elliptic flow of pions in Pb+Pb collisions at $E_{\rm lab}=40A~$GeV with $p_{t} < 2$~GeV/c. UrQMD 
calculations 
are depicted with black lines. The symbols are NA49 data from different analysis methods. The standard method 
(circles), 
cumulant method of order 2 (squares) and cumulant method of order 3 (triangles) are depicted. The 12.5\% most central 
collisions are labeled as central, the centrality 12.5\% -33.5\% as mid-central and 33.5\% -100\% as peripheral. For 
the model calculations the corresponding impact parameters of $b \le 3.4$~fm for central, $b=5-9$~fm for mid-central 
and $b= 9-15$~fm for peripheral collisions have been used.}
\label{figv2ypi40}
\end{figure}

For the elliptic flow of pions (Fig. \ref{figv2ypi40}) in central collisions there are again systematic
uncertainties in the experimental data. Again all three
methods have different shapes and the calculated elliptic flow value is consistent with the measurement with errors. 
The elliptic flow increases to about $1\%$ around midrapidity within the UrQMD model. For
mid-central collisions the picture becomes clearer. The data and the
calculations are in line. Here, the elliptic flow around midrapidity is
about $2-3\%$. Going to peripheral collisions the experimental results
of the standard method show an unsteady behaviour but are compatible with the
UrQMD calculations.

\begin{figure}[t]
\centering
\includegraphics[width=10cm]{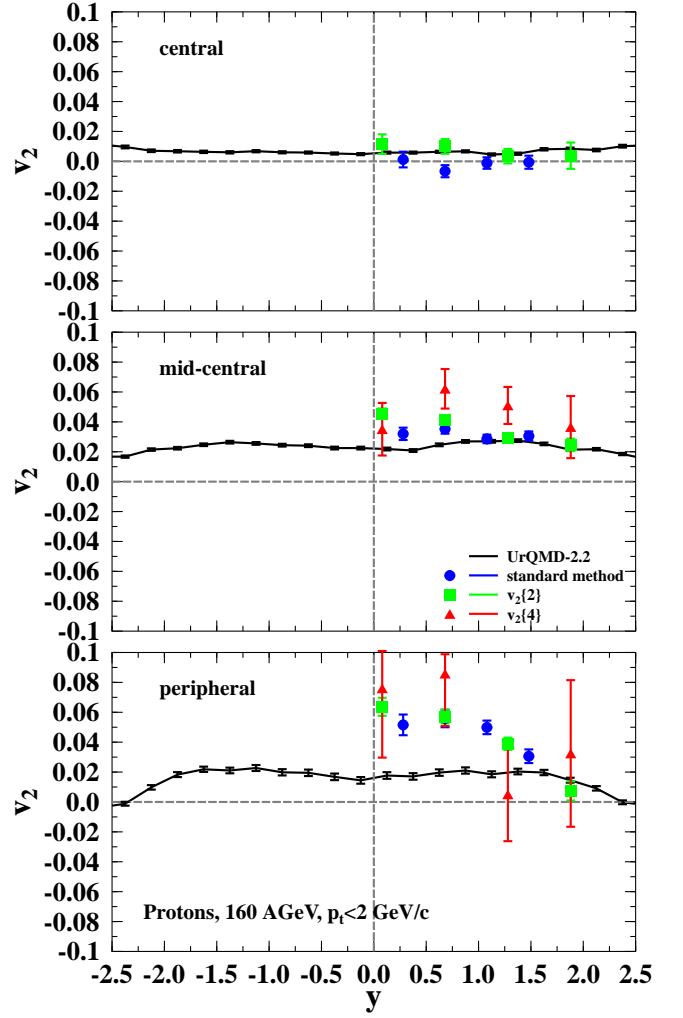}
\caption{(Color online)Elliptic flow of protons in Pb+Pb collisions at $E_{\rm lab}=160A~$GeV with $p_{t} < 2$~GeV/c. UrQMD 
calculations are depicted with black lines. The symbols are NA49 data from different analysis methods. The standard 
method (circles), cumulant method of order 2 (squares) and cumulant method of order 3 (triangles) are depicted. 
The 12.5\% most central collisions are labeled as central, the centrality 12.5\% -33.5\% as mid-central and 
33.5\% -100\% as peripheral. For the model calculations the corresponding impact parameters of $b \le 3.4$~fm for 
central, $b=5-9$~fm for mid-central and $b= 9-15$~fm for peripheral collisions have been used.}
\label{figv2yp160}
\end{figure}

The elliptic flow is expected to be larger in more peripheral collisions
because the anisotropy in coordinate space which is the source of this
flow component is larger. This dependence can be seen very well in the
proton flow at $E_{\rm lab}=160A$~GeV (Fig.\ref{figv2yp160}). In central collisions there is
almost no elliptic flow at all. Going to mid-central collisions the flow
is between $2-6\%$ depending on rapidity and in peripheral collisions even between
$2-8\%$. The model calculations do not show a big difference between
mid-central and peripheral collisions. Both values are about $2\%$ around
midrapidity. The large elliptic flow data in peripheral collisions could be due to the fact of non-flow effects 
and/or large event-by-event $v_2$ fluctuations there, which affect the $v_2$ measurement heavily.

\begin{figure}[t]
\centering
\includegraphics[width=10cm]{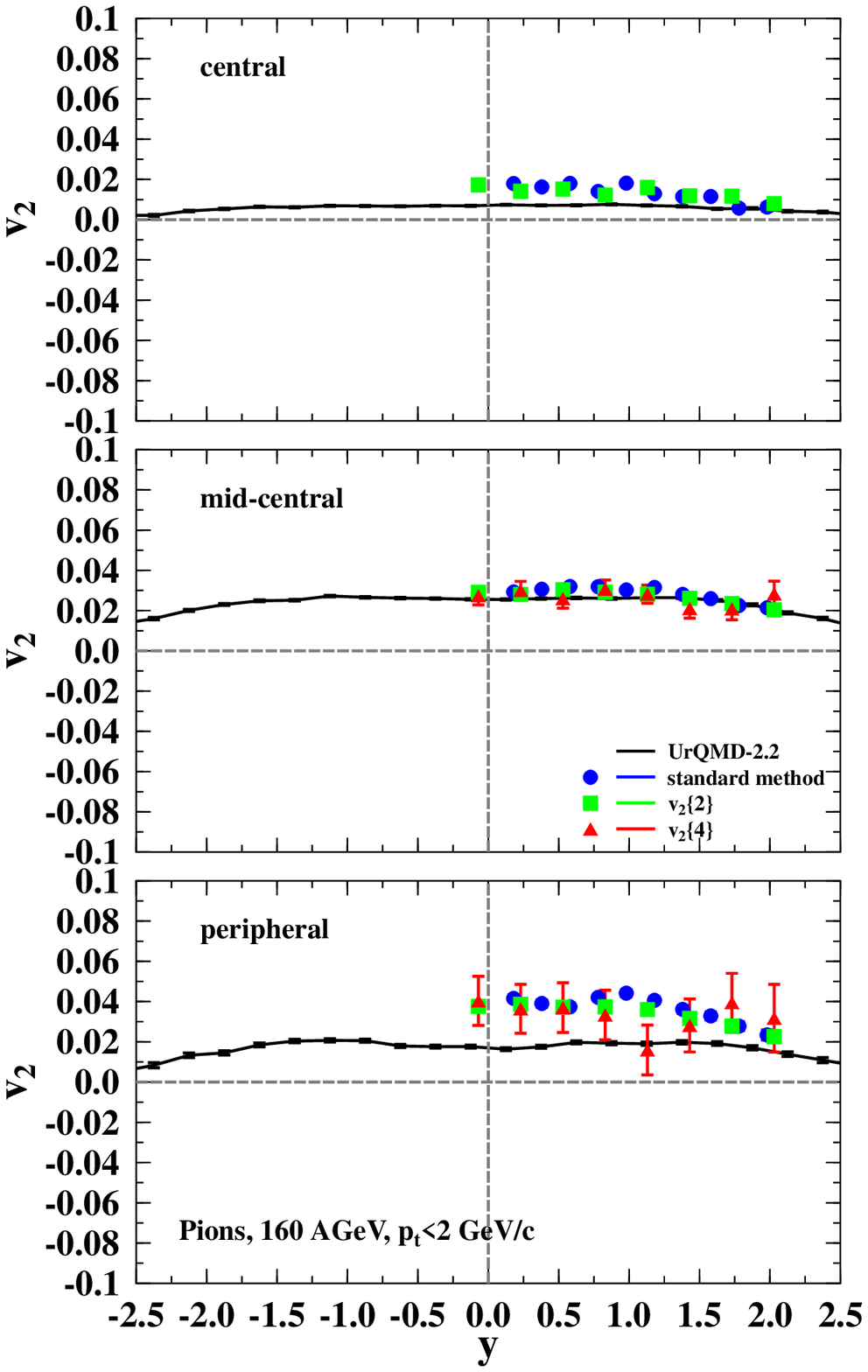}
\caption{(Color online)Elliptic flow of pions in Pb+Pb collisions at $E_{\rm lab}=160A~$GeV with $p_{t} < 2$~GeV/c. UrQMD 
calculations are depicted with black lines. The symbols are NA49 data from different analysis methods. The standard 
method (circles), cumulant method of order 2 (squares) and cumulant method of order 3 (triangles) are depicted. 
The 12.5\% most central collisions are labeled as central, the centrality 12.5\% -33.5\% as mid-central and 
33.5\% -100\% as peripheral. For the model calculations the corresponding impact parameters of $b \le 3.4$~fm for 
central, $b=5-9$~fm for mid-central and $b= 9-15$~fm for peripheral collisions have been used.}
\label{figv2ypi160}
\end{figure}

Fig. \ref{figv2ypi160} shows the elliptic flow of pions at $E_{\rm lab}=160A$~GeV. These newly produced particles show 
almost no elliptic flow in central collisions, i.e. they are emitted isotropically out of the collision region. The
experimental result is between $1-2\%$. In mid-central collisions there
is a very good agreement between the model and data again. For peripheral
collisions the same underestimation of the calculations as for the proton flow in comparison to the data can be seen.

\begin{figure}[t]
\centering
\includegraphics[width=10cm]{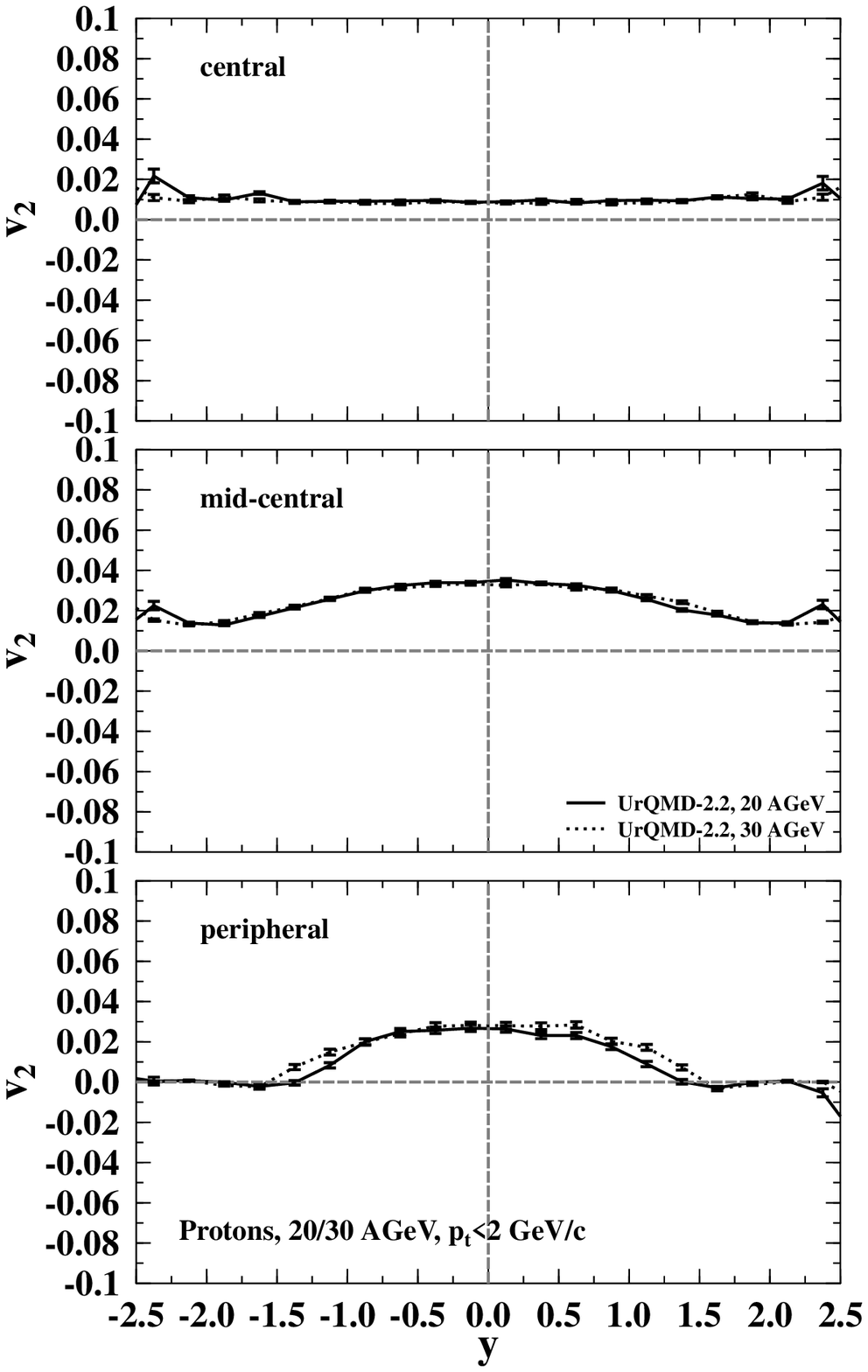}
\caption{Elliptic flow of protons in Pb+Pb collisions at $E_{\rm lab}=20A~$GeV and $E_{\rm lab}=30A~$GeV with 
$p_t < 2$~GeV/c. UrQMD calculations for 20 AGeV are depicted with solid lines while the results for 30 AGeV are 
depicted by dashed lines. Impact parameters of $b \le 3.4$~fm for central, $b=5-9$~fm for mid-central 
and $b= 9-15$~fm for peripheral collisions have been used.}
\label{figv2yp20}
\end{figure}
\begin{figure}[hbt]
\centering
\includegraphics[width=10cm]{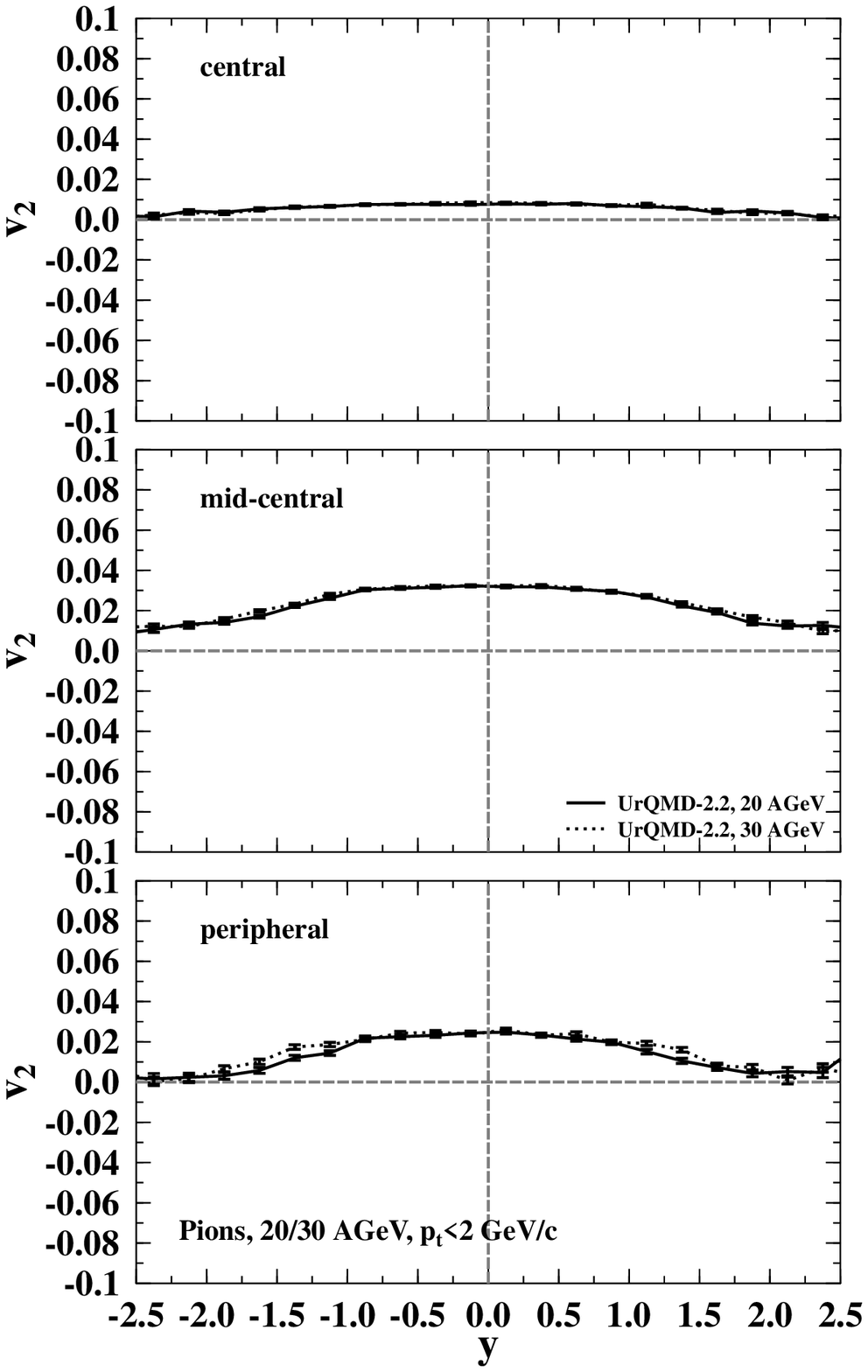}
\caption{Elliptic flow of pions in Pb+Pb collisions at $E_{\rm lab}=20A~$GeV and $E_{\rm lab}=30A~$GeV with 
$p_t < 2$~GeV/c. UrQMD calculations for 20 AGeV are depicted with solid lines while the results for 30 AGeV are 
depicted by dashed lines. Impact parameters of $b \le 3.4$~fm for central, $b=5-9$~fm for mid-central and $b= 9-15$~fm 
for peripheral collisions have been used.}
\label{figv2ypi20}
\end{figure}

Fig. \ref{figv2yp20} and Fig. \ref{figv2ypi20} are predictions for
the rapidity dependence of proton and pion elliptic flow at $E_{\rm lab}=20A$~GeV and $E_{\rm lab}=30A$~GeV. The shown
results are calculated using the UrQMD model with a transverse momentum
cut of $p_t < 2~$GeV/c. The elliptic flow results at this energy of the new
accelerator FAIR at the GSI look rather similar to that at $E_{\rm
lab}=40A$~GeV.

\subsection{Transverse momentum dependence}

\begin{figure}[hbt]
\centering
\includegraphics[width=10cm]{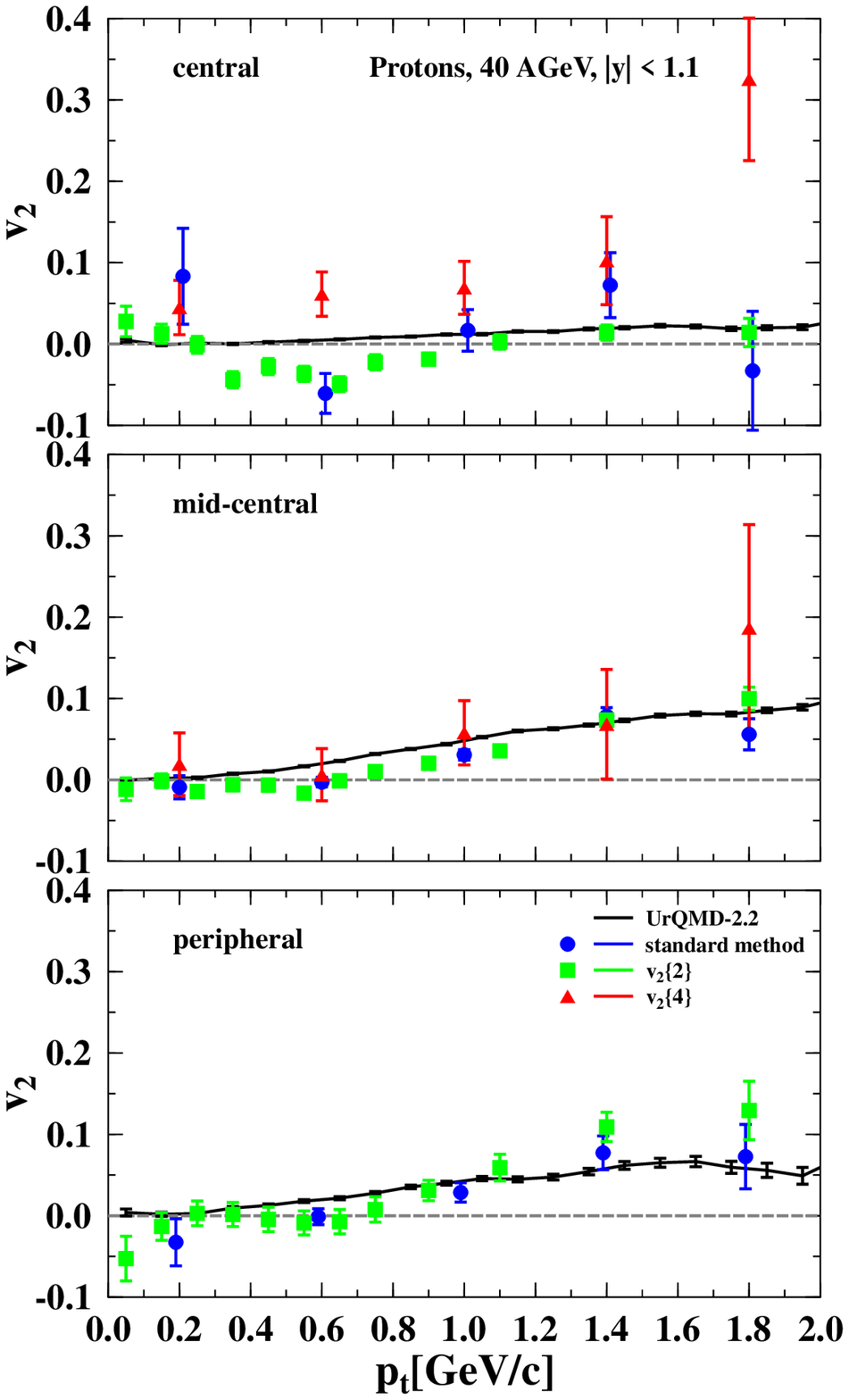}
\caption{(Color online)Elliptic flow of protons in Pb+Pb collisions at $E_{\rm lab}=40A~$GeV with $|y|<1.1$. UrQMD calculations 
are depicted with black lines. The symbols are NA49 data from different analysis methods. The standard method 
(circles), 
cumulant method of order 2 (squares) and cumulant method of order 3 (triangles) are depicted. The 12.5\% most central 
collisions are labeled as central, the centrality 12.5\% -33.5\% as mid-central and 33.5\% -100\% as peripheral. For 
the model calculations the corresponding impact parameters of $b \le 3.4$~fm for central, $b=5-9$~fm for mid-central 
and $b= 9-15$~fm for peripheral collisions have been used.}
\label{figv2ptp40}
\end{figure}
\begin{figure}[hbt]
\centering
\includegraphics[width=10cm]{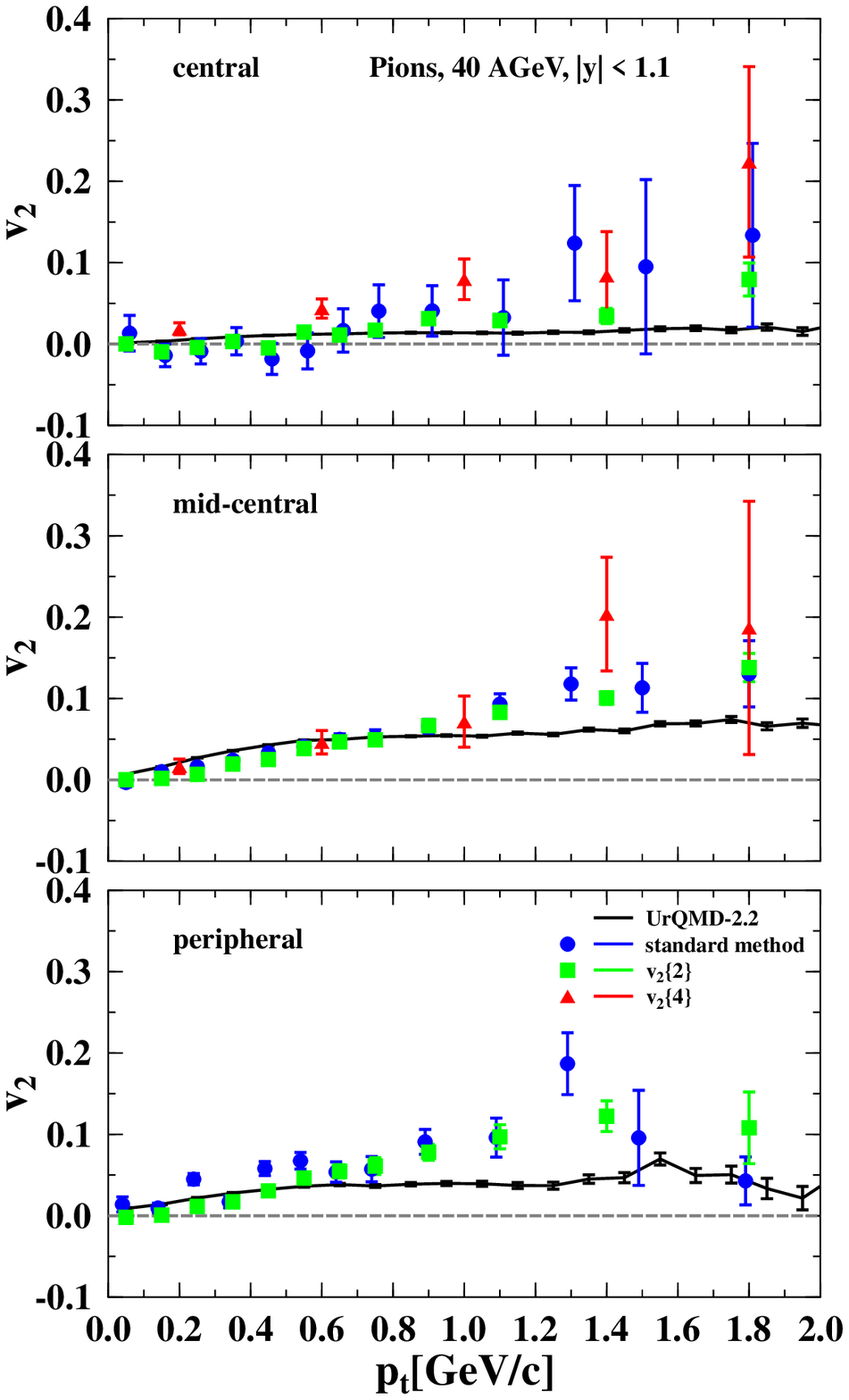}
\caption{(Color online)Elliptic flow of pions in Pb+Pb collisions at $E_{\rm lab}=40A~$GeV with $|y| < 1.1$. UrQMD calculations 
are depicted with black lines. The symbols are NA49 data from different analysis methods. The standard method 
(circles),
 cumulant method of order 2 (squares) and cumulant method of order 3 (triangles) are depicted. The 12.5\% most central 
 collisions are labeled as central, the centrality 12.5\% -33.5\% as mid-central and 33.5\% -100\% as peripheral. For 
 the model calculations the corresponding impact parameters of $b \le 3.4$~fm for central, $b=5-9$~fm for mid-central 
 and $b= 9-15$~fm for peripheral collisions have been used.}
\label{figv2ptpi40}
\end{figure}
\begin{figure}[hbt]
\centering
\includegraphics[width=10cm]{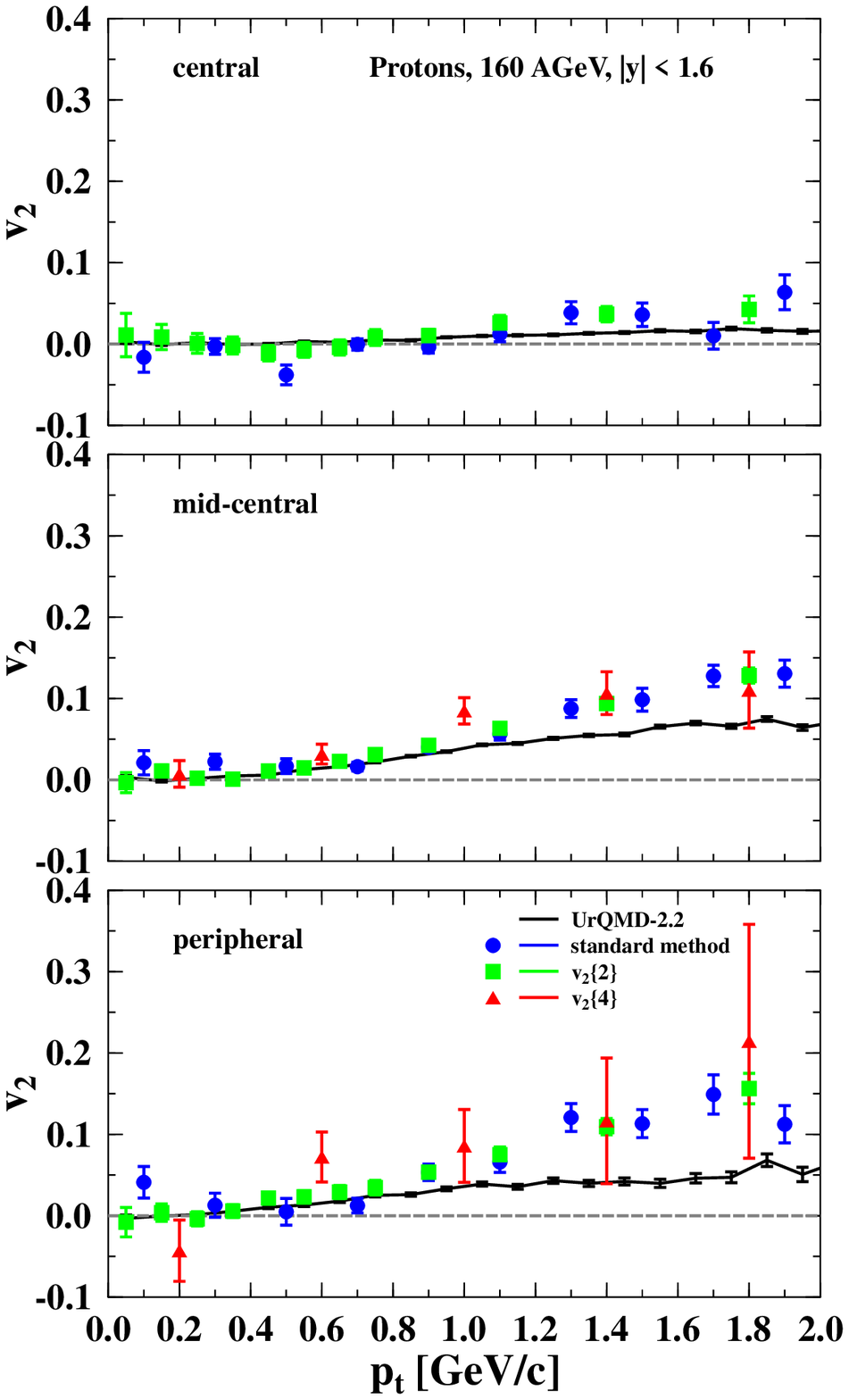}
\caption{(Color online)Elliptic flow of protons in Pb+Pb collisions at $E_{\rm lab}=160A~$GeV with $|y| < 1.6$. UrQMD calculations 
are depicted with black lines. The symbols are NA49 data from different analysis methods. The standard method 
(circles), 
cumulant method of order 2 (squares) and cumulant method of order 3 (triangles) are depicted. The 12.5\% most central 
collisions are labeled as central, the centrality 12.5\% -33.5\% as mid-central and 33.5\% -100\% as peripheral. For 
the model calculations the corresponding impact parameters of $b \le 3.4$~fm for central, $b=5-9$~fm for mid-central 
and $b= 9-15$~fm for peripheral collisions have been used.}
\label{figv2ptp160}
\end{figure}

The elliptic flow of protons at $E_{\rm lab}=160A$~GeV (Fig. \ref{figv2ptp160}) in central
collisions looks similar to that at $40A$~GeV. There is a smooth increase
from zero at $p_t=0$ to about 2\% at $p_t=2~$GeV/c. The underestimation of
the flow by the transport model calculations at higher energies is
visible in the results for mid-central and peripheral collisions at
high $p_t$. At low $p_t$ the calculations are in line with the data.
\begin{figure}[hbt]
\centering
\includegraphics[width=10cm]{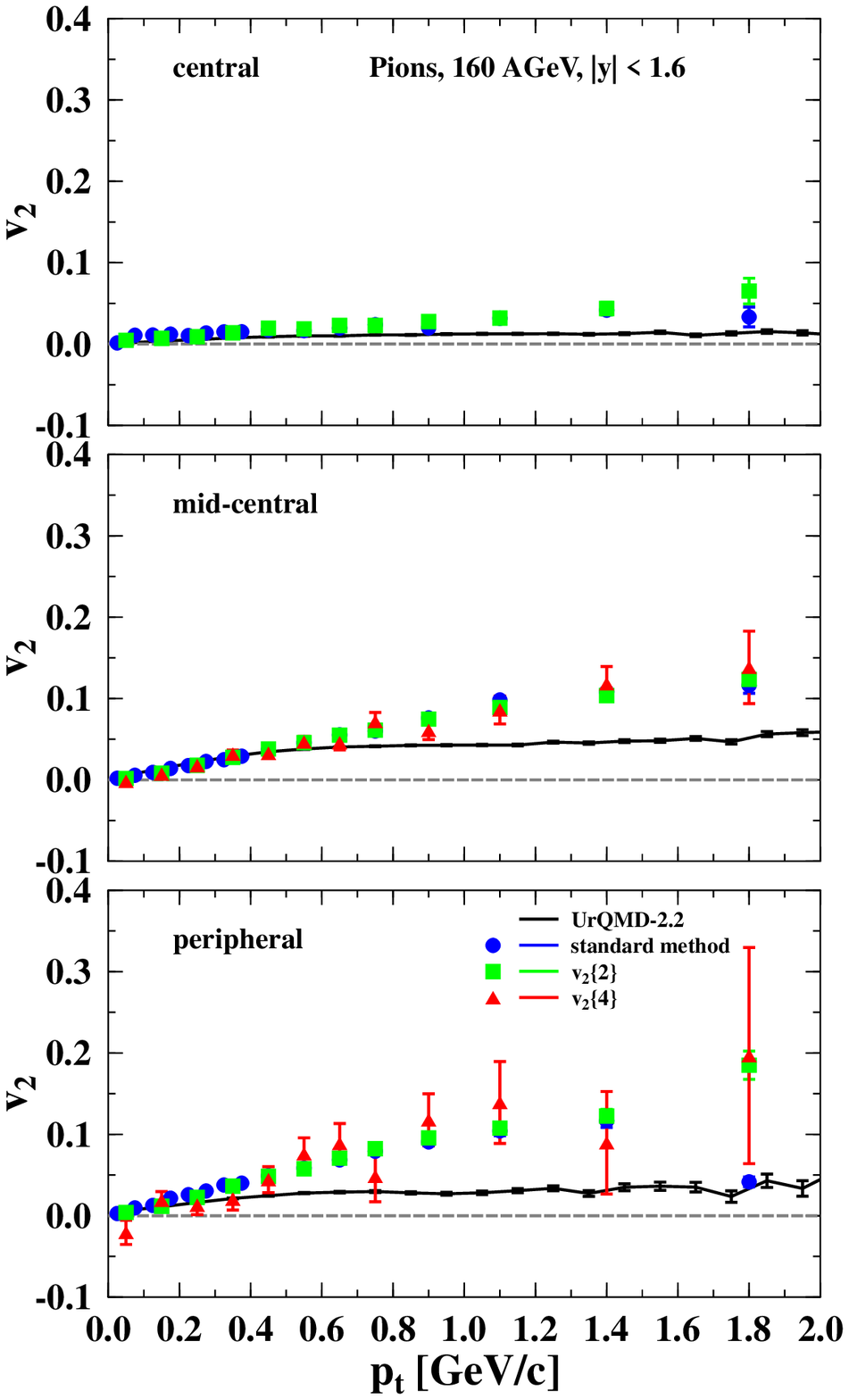}
\caption{(Color online)Elliptic flow of pions in Pb+Pb collisions at $E_{\rm lab}=160A~$GeV with $|y| < 1.6$. UrQMD calculations 
are depicted with black lines. The symbols are NA49 data from different analysis methods. The standard method 
(circles),
 cumulant method of order 2 (squares) and cumulant method of order 3 (triangles) are depicted. The 12.5\% most central 
 collisions are labeled as central, the centrality 12.5\% -33.5\% as mid-central and 33.5\% -100\% as peripheral. For 
 the model calculations the corresponding impact parameters of $b \le 3.4$~fm for central, $b=5-9$~fm for mid-central 
 and $b= 9-15$~fm for peripheral collisions have been used.}
\label{figv2ptpi160}
\end{figure}

For the elliptic flow of protons as a function of transverse momentum there is a
different rapidity cut used in the data than in the calculations. For
proton and pion flow at $E_{\rm lab}=40A$~GeV it is $-0.1 < y < 1.1$ and
for $E_{\rm lab}=160A$~GeV it is $0.1 < y < 1.6 $ for the cumulant order
measurements. For the standard reaction plane method data, larger rapidity bins have been
used for the integration. To improve the statistics of the UrQMD results we use symmetric rapidity cuts as $|y| < 1.1$ 
for $40A$~GeV and $|y| < 1.6$ for $160A$~GeV. Since the elliptic flow is
symmetric in rapidity this can be done without problems. The elliptic
flow of protons at $E_{\rm lab}=40A$~GeV (Fig. \ref{figv2ptp40}) increases only to about $3\%$ for central and 
peripheral collisions in the calculated results. For mid-central collisions there
is a steady increase with the increase of transverse momentum to about $10$\%
which is in line with the data.

For the calculated pion elliptic flow at $E_{\rm lab}=40A$~GeV (Fig. \ref{figv2ptpi40}) a quite
similar behaviour as for protons is seen. At high $p_t$ the
experimental results have large statistical error bars. At low $p_t$
the experimental is not plagued by statistical and systematic errors and one observes good agreement between model and
data. There is a increase of elliptic flow with increasing transverse
momentum because the larger the momentum of the particle the earlier it escapes the collision zone \cite{Bass:1994af}. These high energy 
particles carry the signal of the very early stage of the collision where the coordinate space asymmetry is most 
pronounced \cite{Lu:2006qn}.

For pion elliptic flow at 160 AGeV the picture is very similar to the
plot of the proton flow at this energy.

\begin{figure}[hbt]
\centering
\includegraphics[width=10cm]{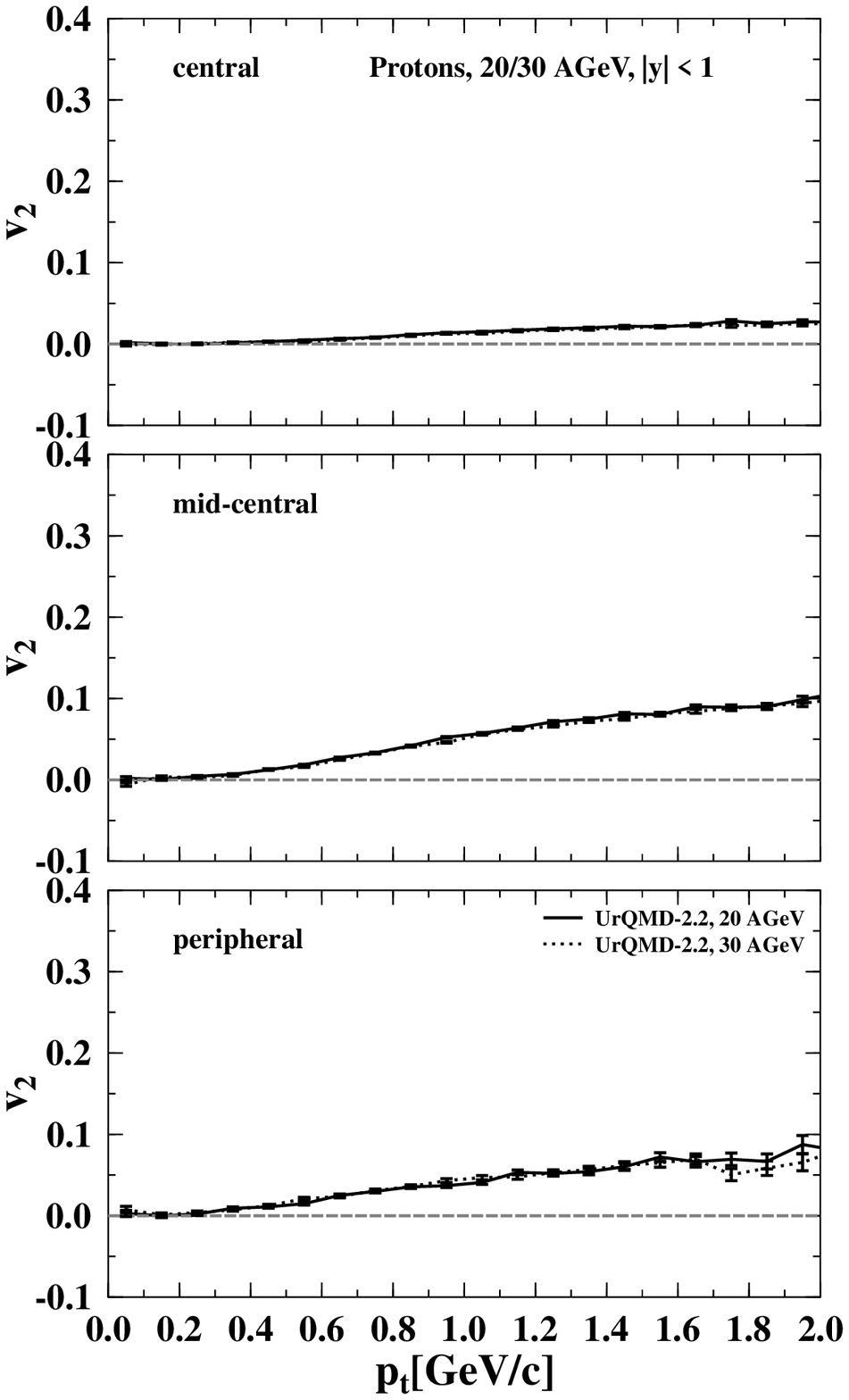}
\caption{Elliptic flow of protons in Pb+Pb collisions at $E_{\rm lab}=20A~$GeV and $E_{\rm lab}=30A~$GeV with 
$|y| < 1$. UrQMD calculations for 20 AGeV are depicted with solid lines while the results for 30 AGeV are depicted 
by dashed lines. Impact parameters of $b \le 3.4$~fm for central, $b=5-9$~fm for mid-central and $b= 9-15$~fm for 
peripheral collisions have been used.}
\label{figv2ptp20}
\end{figure}

The predictions for the transverse momentum dependence of proton elliptic flow
at $E_{\rm lab}=20A$~GeV and $E_{\rm lab}=30A$~GeV (Fig. \ref{figv2ptp20}) have mostly the same shape as the 
calculations
at $40A$~GeV.  
\begin{figure}[hbt]
\centering
\includegraphics[width=10cm]{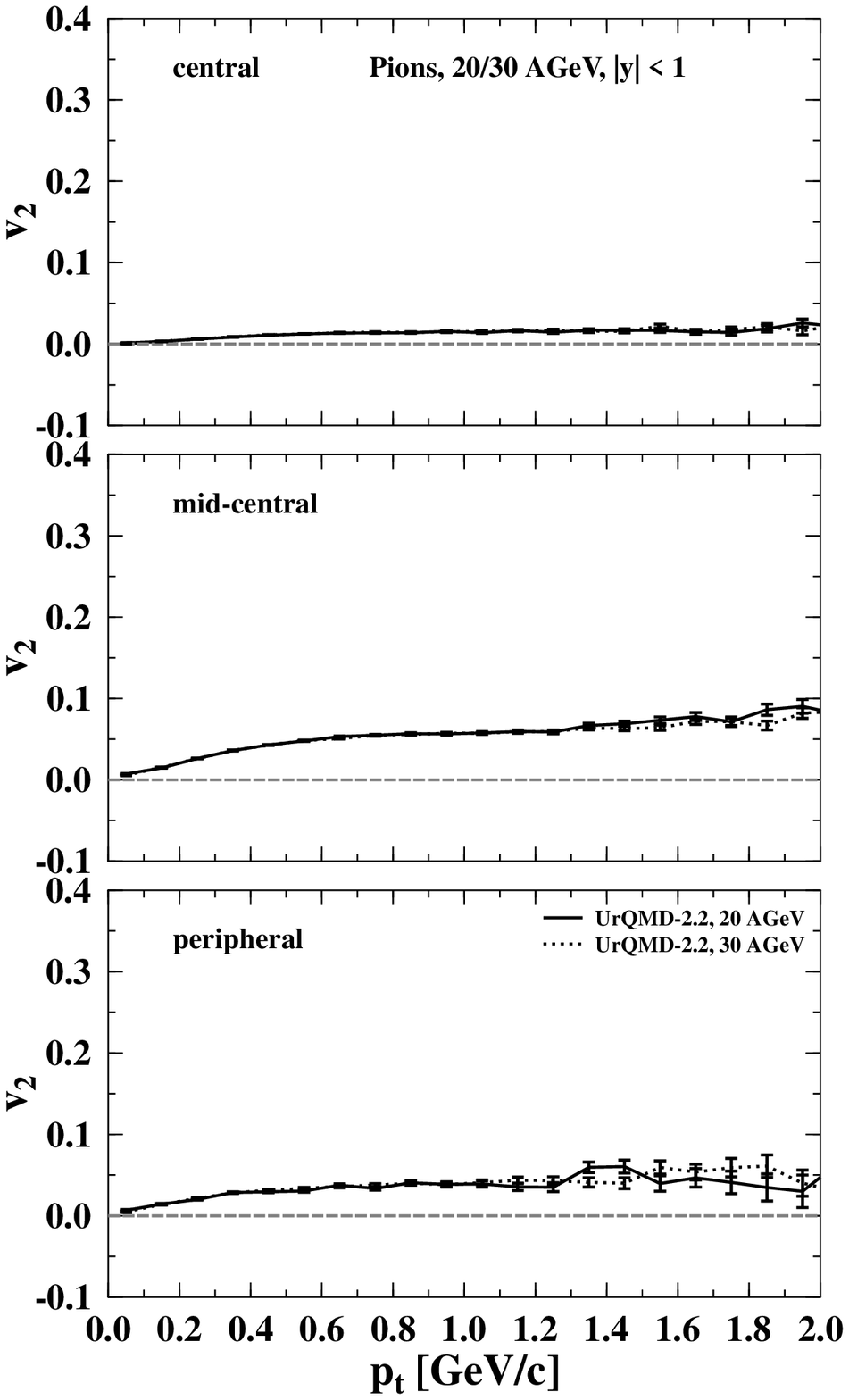}
\caption{Elliptic flow of pions in Pb+Pb collisions at $E_{\rm lab}=20A~$GeV and $E_{\rm lab}=30A~$GeV with $|y| < 1$. 
UrQMD calculations for 20 AGeV are depicted with solid lines while the results for 30 AGeV are depicted 
by dashed lines. Impact parameters of $b \le 3.4$~fm for central, $b=5-9$~fm for mid-central and $b= 9-15$~fm for 
peripheral collisions have been used.}
\label{figv2ptpi20}
\end{figure}

The elliptic flow of pions at $20A$~GeV and $30A$~GeV shown in Fig. \ref{figv2ptpi20} are also similar to the
calculations for the pion elliptic flow at $40A$~GeV. But there is a
difference for peripheral collisions. At the higher energy there is a
lower value than for the lower energy.

\subsection{Excitation function}
\begin{figure}[hbt]
\centering
\includegraphics[width=10cm]{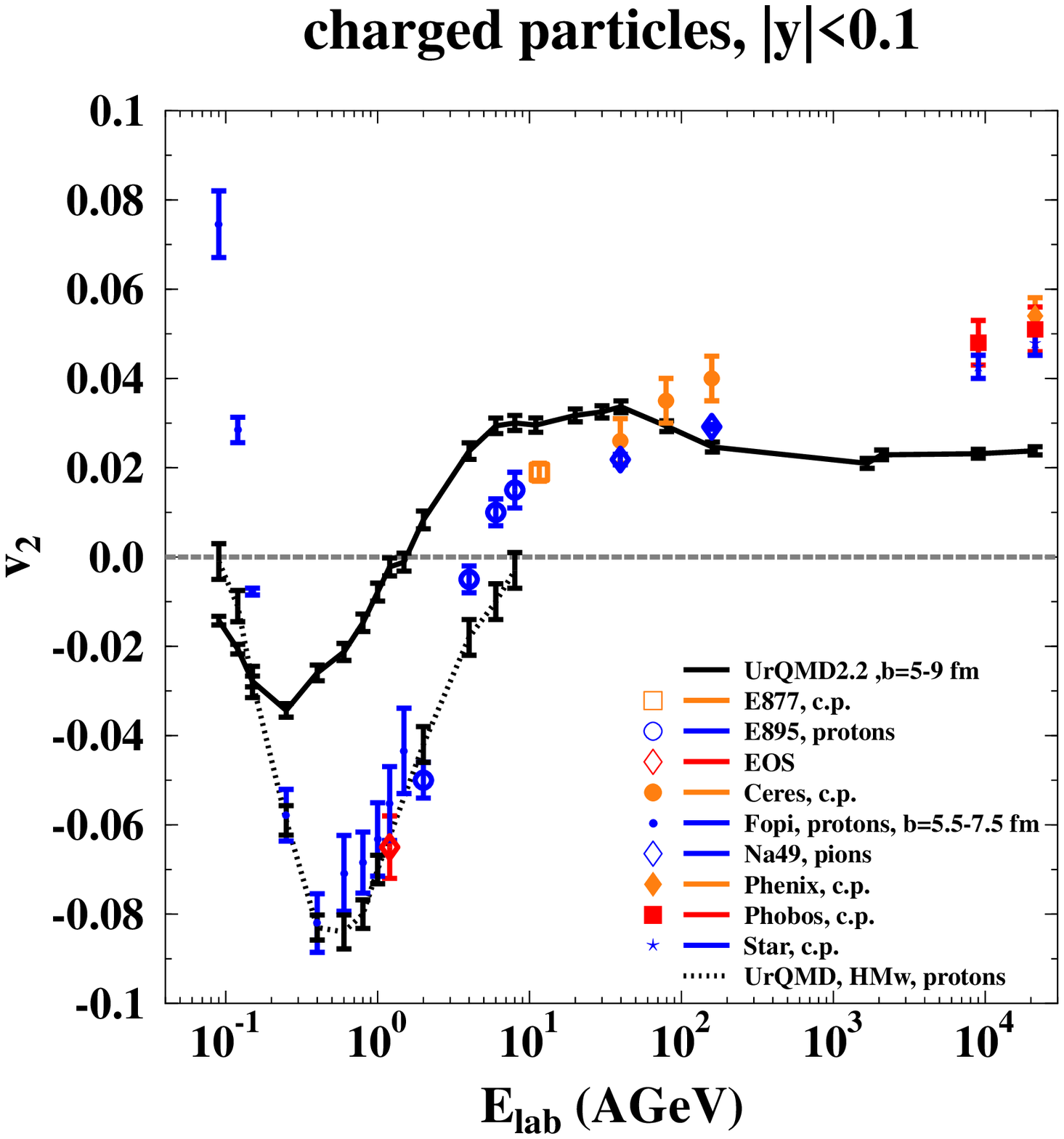}
\caption{(Color online) The calculated energy excitation function of elliptic flow of charged particles in Au+Au/Pb+Pb 
collisions in mid-central collisions (b=5-9 fm)with $|y|<0.1$(full line). This curve is compared to data from 
different experiments for mid-central collisions. For E895 \cite{Pinkenburg:1999nv,Chung:2001qr} and 
FOPI \cite{Andronic:2004cp} there is the elliptic flow of protons and for NA49 \cite{Alt:2003ab} it is the elliptic 
flow of pions. For E877, CERES \cite{Filimonov:2001fd,Slivova:2002wj,CERES:YUN02}, PHENIX \cite{Esumi:2002vy},
PHOBOS\cite{Manly:2002uq} and STAR \cite{Ray:2002md}there is data for the charged particle flow. The dotted line 
in the low energy regime depicts UrQMD calculations with the mean field \cite{Li:2006ez}.}
\label{figv2chexc}
\end{figure}

The excitation function of charged particle elliptic flow is
compared to data over a wide energy range (Fig. \ref{figv2chexc}), i.e from $E_{\rm lab}=90~A$MeV to 
$\sqrt{s_{NN}}=200$~GeV. The squeeze-out effect at low energies and the change to in-plane emission at higher 
energies is nicely observed in the excitation function. The symbols indicate the data for charged particles 
from different experiments. Note however, that in the low energy regime
there are only experimental data points for protons. For beam energies below
2 AGeV most of the charged particles are also protons because there is not
enough energy to produce many new particles. Going to higher energies the
elliptic flow of pions and charged particles are very similar. The
rapidity cut of $|y| < 0.1$ has been used for the whole energy range
despite the fact that the data at higher energies is within $|y| < 0.5$. This has been done
to avoid too much changes in the parameters and this choice gives reasonable results over the whole energy range. 
We have checked that the results at higher energies are not affected by the choice of this narrower rapidity window. 

\begin{figure}[hbt]
\centering
\includegraphics[width=10cm]{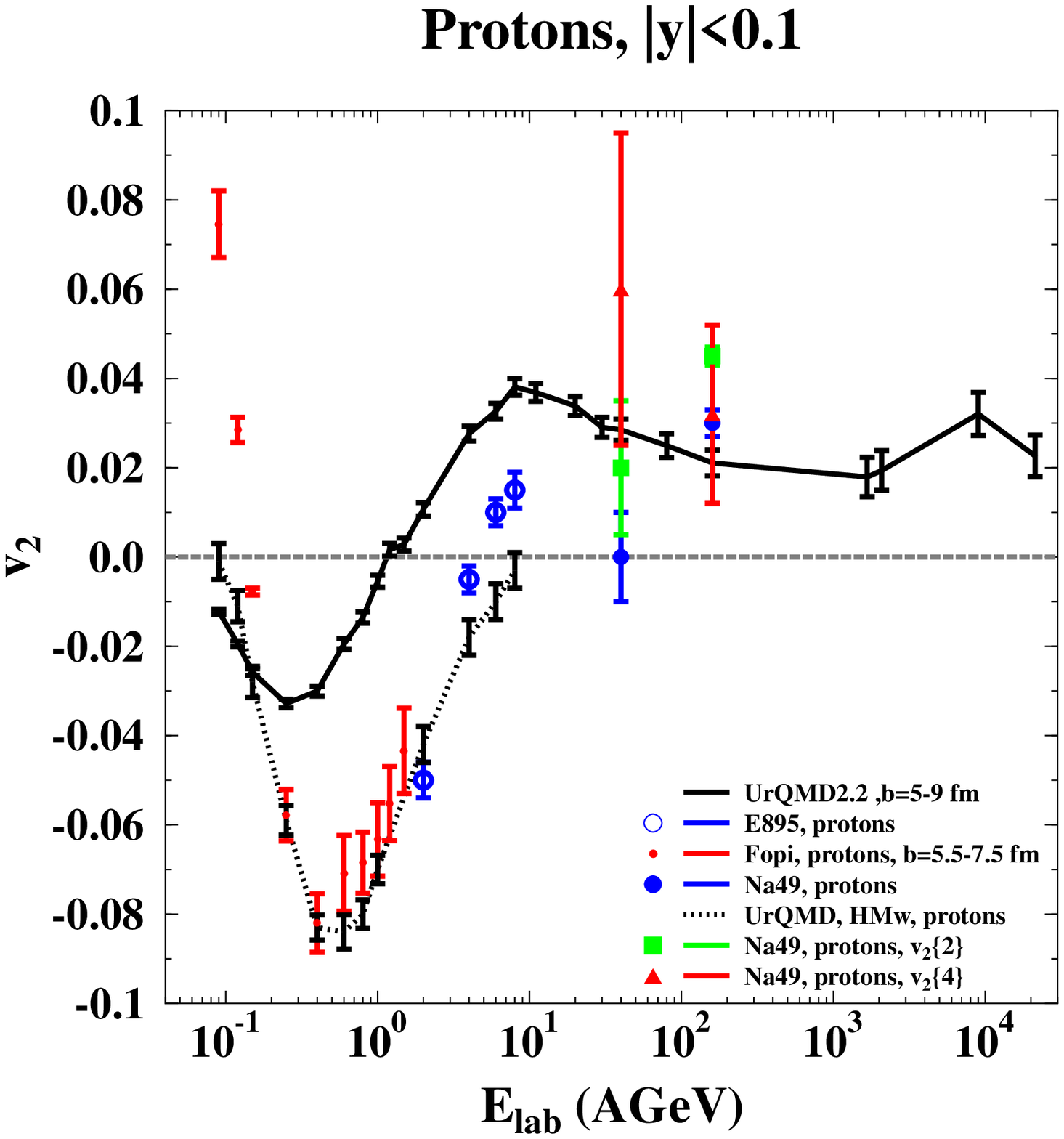}
\caption{(Color online) The calculated energy excitation function of elliptic flow of protons in Au+Au/Pb+Pb 
collisions in mid-central collisions (b=5-9 fm)with $|y|<0.1$(full line). This curve is compared to data from 
different experiments for mid-central collisions. For E895 \cite{Pinkenburg:1999nv}\cite{Chung:2001qr}, FOPI 
\cite{Andronic:2004cp} and NA49 \cite{Alt:2003ab} there is the elliptic flow of protons. The dotted line in the 
low energy regime depicts UrQMD calculations with included nuclear potential.}
\label{figv2nuy0.1}
\end{figure}

At low energies $E_{beam}\sim 0.1 - 6$  A GeV the
squeeze-out effect, i.e. the elliptic flow out-of-plane, is clearly seen
in the data as well as in the calculations, especially when the mean field is considered. At such energies, it is well 
known that both the mean field and the two-body collision are equally important  to reproduce quantitatively the 
experimental results \cite{Danielewicz:1999vh,Danielewicz:1998vz,Pan:1992ef}. In this paper we adopt a hard equation 
of state  with momentum dependence (HM-EoS) which 
was updated recently in UrQMD model \cite{Li:2005gf}. Meanwhile, the two-body scattering in heavy ion collisions 
might be modified by the nuclear medium. In order to consider (partly) the medium effect, the nucleon-nucleon elastic 
scattering cross sections are modified to depend on the nuclear density, the isospin-asymmetry and the two-nucleon 
relative momentum, besides the center-of-mass energy of two nucleons. This treatment was investigated based on the 
relativistic Dirac-Brueckner-Hartree-Fock (DBHF) theory as well as the relativistic mean field (RMF) theory, please 
see Ref.\  \cite{Li:2006ez} for details. Here we show the calculation results with the HM-EoS and with the 
DBHF-like medium modification on nucleon-nulceon elastic cross sections (HMw)  \cite{Li:2006ez}. 

In the SPS regime the model calculations are quite in line with
the data, especially with the NA49 results. Above $E_{\rm lab}=160A~$GeV the calculation underestimates the elliptic 
flow. At the highest RHIC energy there are about 5\% flow in the data while the
model calculation provides only half of this value. This can be explained by
assuming a lack of pressure in the transport model at these energies. 

In Fig. \ref{figv2nuy0.1} the excitation function of elliptic flow of nucleons is presented. The rapidity cut of 
$|y|<0.1$ has been used because this is the
appropriate one to compare with the data at lower energies. Due to this cut now also the calculation in the cascade 
mode (without
nuclear potential) reaches negative values at low energies. With
included potential the model is in line with experimental data. The
UrQMD result in the SPS energy range lies in between the NA49 measurement of
elliptic flow of protons. These values have been extracted from the
differential plots of elliptic flow over rapidity for mid-central
collisions discussed above in this paper.

The observed proton flow $v_2$ below $\sim$ $5A~$GeV is smaller than
zero, which corresponds to the squeeze-out predicted by hydrodynamics
long ago
\cite{Hofmann74,Hofmann:1976dy,Stoecker:1980vf,Stocker81,Stoecker:1981pg,Stoecker:1986ci}. At
higher energies, $10-160 A~$GeV, an increase of the flow
$v_2$ to a maximum around $E_{\rm lab}=10A~$GeV followed by a decrease to about 2 \% and a saturation is 
predicted from the string-hadronic transport model.  In fact,
the $158A~$GeV data of the NA49 Collaboration suggest that a
smooth increase proceeds between AGS and SPS.

The ``collapse'' of $v_2$ (strong negative value) for protons around midrapidity at $40A~$GeV
is only pronounced in the standard method data. The
UrQMD calculations, without a phase transition, show a robust ~3\% flow of protons. One cannot say anything 
about a clear underestimation at high energies in this case because integrated proton flow data at RHIC is still not 
available.  
 
\section{Summary}

We have compared UrQMD calculations to recent NA49 data. In general, a good agreement between data and calculation 
is found. There seem to be systematic uncertainties in the measurement method looking at the different results. For 
example for the directed flow data the effect of momentum conservation on the flow data can be seen. The slope around 
midrapidity of the  rapidity distributions of proton directed flow becomes negative around $E_{\rm lab}= 40 A~$GeV. This 
cannot be reproduced by the transport model calculations.   
The excitation function of elliptic flow shows strong negative flow at low energies - the ``squeeze-out''-effect - 
which can quantitatively only be reproduced by including a nuclear potential in the calculation. At high energies we 
observed an underestimation of the elliptic flow of charged particles in the present model. This can possibly be 
explained by assuming a lack of pressure in the early stage of the collisions at high energies. 
It will be very interesting to see what happens in the lower energy regime ($20/30 A ~$GeV) when high quality CBM-FAIR 
data becomes accessible.   

\begin{acknowledgments}
We are grateful to the Center for the Scientific Computing (CSC) at Frankfurt for the computing resources. The
authors thank Alexander Wetzler for helpful and stimulating discussions. Q. Li thanks the Alexander von 
Humboldt-Stiftung for
a fellowship. This work was supported by GSI and BMBF. 
\end{acknowledgments}

\end{document}